\documentclass[a4paper,11pt]{article}
\pdfoutput=1 

\usepackage{jinstpub} 

\usepackage{soul}
\usepackage{color}
\usepackage[dvipsnames]{xcolor}
\usepackage{multicol} 
\usepackage{array} 
\usepackage{wasysym}
\usepackage{hyperref}
\usepackage{eurosym}
\usepackage{pifont}
\usepackage{subfigure}
\usepackage[bottom]{footmisc}
\usepackage{gensymb}

\usepackage{multirow}
\usepackage{comment}
\usepackage{xspace}
\usepackage{siunitx}

\newcommand{\three}{\ensuremath{3\times1\times1\mbox{ m}^3}\xspace}

\title{Study of scintillation light collection, production and propagation in a 4 tonne dual-phase LArTPC}

\author[a]{B.\,Aimard}
\author[p]{L.\,Aizawa}
\author[v]{C.\,Alt}
\author[b]{J.\,Asaadi}
\author[e]{M.\,Auger}
\author[m]{V.\,Aushev}
\author[t]{D.\,Autiero}
\author[g]{A.\,Balaceanu}
\author[a]{G.\,Balik}
\author[t]{L.\,Balleyguier}
% ...

\affiliation[a]{LAPP, Universit\'e Savoie Mont Blanc, CNRS/IN2P3, Annecy, France}
\affiliation[b]{University of Texas Arlington, Arlington, USA}
% ...
\affiliation[e]{University of Bern, Albert Einstein Center for Fundamental Physics, Laboratory for High Energy Physics (LHEP), Bern, Switzerland}
% ...
\affiliation[g]{Horia Hulubei National Institute for R\&D in Physics and Nuclear Engineering - IFIN-HH, Bucharest - Magurele, Romania}
% ...
\affiliation[m]{Kyiv National University, Kyiv, Ukraine}
% ...
\affiliation[p]{Iwate University, Department of Electrical Engineering and Computer Science, Morioka, Iwate, Japan}
% ...
\affiliation[t]{Institut de Physique des 2 Infinis (IP2I), CNRS/IN2P3, Universit\'e de Lyon, Universit\'e Claude Bernard Lyon 1, Villeurbanne, France}
% ...
\affiliation[v]{ETH Zurich, Institute for Particle Physics, Zurich, Switzerland}

\emailAdd{laura.zambelli@lapp.in2p3.fr}

\abstract{The \three demonstrator is a dual phase liquid argon time projection chamber that has recorded cosmic rays events in 2017 at CERN. The light signal in these detectors is crucial to provide precise timing capabilities. The performance of the photon detection system, composed of five PMTs, are discussed. The collected scintillation and electroluminescence light created by passing particles has been studied in various detector conditions. In particular, the scintillation light production and propagation processes have been analyzed and compared to simulations, improving the understanding of some liquid argon properties.}


\begin{document}
\maketitle
\flushbottom
\section*{Introduction}
Liquid argon (LAr) is widely used as active medium for neutrino and dark matter detection. They are studied through their interaction with argon nuclei generating secondary particles which can potentially be detected.
In particular, charged particles and gamma rays excite and ionize liquid argon, leading to the production of scintillation light and ionization charge. 
The time projection chamber (TPC) is based on the collection of these two signals to fully reconstruct the event. The charge signal allows a 3D imaging with a mm$^3$ resolution scale, the scintillation light signal provides the absolute timing of each reconstructed tracks. %
By using the imaging and calorimetric capabilities of liquid argon, the TPC has an excellent e/$\gamma$ separation power which makes this technology very promising for the future of neutrino oscillation experiments.

The Deep  Underground  Neutrino  Experiment (DUNE) envisages deploying four \SI{17}{\kilo\tonne} LArTPC modules (\SI{10}{\kilo\tonne} active mass) in the Sanford Underground Research Facility in South Dakota, USA.  Using a powerful neutrino beam originating from Fermilab, DUNE aims to address key questions in neutrino physics and astroparticle physics~\cite{Abi:2020wmh, Abi:2020evt, Abi:2020oxb, Abi:2020loh, duneIDRv3}. 
The DUNE physics program includes precision measurements of the parameters that govern neutrino oscillations with the goal of unveiling the leptonic CP violating phase and the neutrino mass ordering within 10 years of operation. Given the detector size and capabilities, the physics case of DUNE expands to nucleon decay searches and the possible detection of the electron neutrino flux from a core-collapse supernova within our galaxy. 

In order to gain experience in building and operating such large-scale LArTPC detectors, an R\&D program is currently underway at the CERN Neutrino Platform.  Two prototypes have been constructed with the specific aim of testing the design, assembly, installation procedures, detector operations, as well as data acquisition, storage, processing, and analysis.  They both have similar sizes, each with approximately \SI{300}{\tonne} of active mass, and explore two possible LArTPC designs foreseen for the DUNE far detector modules.  While the first one consists of argon in liquid state only (ProtoDUNE Single-Phase~\cite{tdr_protoduneSP}), the second adds a layer of gaseous argon to enable charge amplification before collection (ProtoDUNE Dual-Phase~\cite{wa105, protoDUNElight}). As a consequence, the dual phase (DP) technology can lower the energy detection threshold with a high signal-to-noise ratio over drift distances exceeding \SI{10}{\meter}.

Prior to the construction of ProtoDUNE-DP, a demonstrator with an active volume of \three (\SI{4.2}{\tonne}) has been built at CERN~\cite{311} in 2016 and recorded cosmic rays in 2017. The demonstrator allowed to validate the concept of a non-evacuated industrial cryostat, test several key sub-systems for ProtoDUNE-DP, and demonstrate for the first time the capabilities of the dual-phase LArTPC technology at the tonne scale.

The study of the scintillation light collected in the \three dual-phase LArTPC demonstrator is the topic of this paper. In the first section, our current understanding and unknowns on the scintillation light production and propagation mechanisms in LAr are reviewed. The detector design is then detailed in Section~\ref{sec2}. The third section presents the performance of our light detector sensors. The collected data is summarized in Section~\ref{sec4}.
The light simulation tools are explained in Section~\ref{sec5}. The analyses of scintillation light propagation and production are discussed in Sections~\ref{sec6} and \ref{sec7}, respectively. Finally, studies using the electroluminescence light signal, emitted during charge amplification in gas, are summarized in the last section.
\section{Scintillation light in a dual-phase LArTPC}
\label{sec1}

The detection of scintillation light in LArTPC detectors is a key feature of this technology. This signal provides the absolute timing of recorded tracks, which is necessary for the full 3D reconstruction. Moreover, the scintillation can be used as an internal self trigger for rare phenomena, and may improve the detector calorimetry performance. For these reasons, a fundamental understanding of the scintillation light production and propagation properties is required. 

In LAr, the scintillation light is generated by the decay to ground state of diatomic excited molecules. These excimers can be either generated directly through argon excitation, or through the recombination of ionized argon with an electron~\cite{Kubota,Amoruso}:

\begin{eqnarray*}
\mathrm{Excitation}&:& \; Ar^* + Ar \rightarrow Ar_2^* \rightarrow Ar + Ar + \gamma  \\
\mathrm{Recombination}&:& \; Ar^+ + Ar \rightarrow Ar_2^+ e^- \rightarrow Ar_2^* \rightarrow Ar + Ar + \gamma
\end{eqnarray*}
In both processes, the photons are emitted isotropically with a wavelength peaked at \SI{128}{\nano\meter} in the VUV region~\cite{Heindl:2010zz}, and together constitute the primary scintillation light (S1) signal. In the absence of drift field, the light yield is of the order of $5\times 10^4$~photons/MeV.

The recombination process requires an electron cloud surrounding the $Ar^+$ to occur. Hence, the scintillation yield due to recombination is dependent on both the energy loss (dE/dx) of the particle interacting in the TPC and the electric drift field~\cite{Cennini:1999ih}. As a consequence, the charge production is anti-correlated with this light production process.

Depending on the spin orientations, the excimer can be either produced in a singlet or in a triplet energy state.  While the singlet state has a lifetime of $\sim$\SI{6}{\nano\second}, the lifetime of the triplet state is $\sim$\SI{1.6}{\micro\second}~\cite{Hitachi}.  
The fast rise of the scintillation light from the singlet state is the feature used for event timing and internal trigger. 
However, the time structure of the scintillation light contains much more information which can be exploited with a suitable photon detection system. The ratio of singlet to triplet state depends on the ionization density; this feature can be used for particle identification~\cite{Lippincott}.
When the LAr contains trace contaminants, the excitation energy of the excimers may be transferred to certain molecules like O$_2$ and N$_2$ instead of being released by the emission of a photon.
The measurement of the triplet state lifetime is then a handle to estimate the level of impurities~\cite{Acciarri}.

While primary scintillation occurs in all LAr detectors, secondary scintillation (S2) or electroluminescence light is produced in detectors based on the dual phase concept.  In this case, electrons produced in LAr are drifted to the top and extracted to an ultra pure gas layer by means of a strong local electric field.  In the gas phase, electrons are accelerated with high electric fields in order to generate Townsend avalanches. During this process, argon excimers are produced and emit photons in the same VUV region as the primary scintillation light during their decay. The S2 signal is a reflection of the charge signal: its duration, start and end times are directly connected to the track topology. The time difference between the prompt S1 and the delayed S2 signals relates to the drift distance of the electrons.

Concerning the propagation of the scintillation light in LAr, two processes are relevant: absorption and Rayleigh scattering.  While pure LAr is highly transparent to its own scintillation light, even tiny amounts of methane~\cite{Jones:2013mfa}, nitrogen~\cite{Jones:2013bca}
and oxygen~\cite{Acciarri} impurities can lead to significant light absorption and will limit the low energy detection threshold. On the other hand, the Rayleigh scattering is the process of light elastically scattered off particles smaller than its wavelength. 
The Rayleigh scattering increases the travelling distance of photons between their production point to their detection and therefore spreads the arrival time of the photons.  Literature values of the Rayleigh scattering length in LAr for VUV photons lay between 55 and \SI{90}{\centi\meter} (theory)~\cite{Grace, Seidel} and 66 and \SI{163}{\centi\meter} (experimental)~\cite{Ishida,Neumeier}.  A detailed review about liquid noble detectors, including scintillation light, can be found in~\cite{Chepel}.

\section{The \three demonstrator}
\label{sec2}

\begin{figure}
    \centering
    \includegraphics[width=0.8\textwidth]{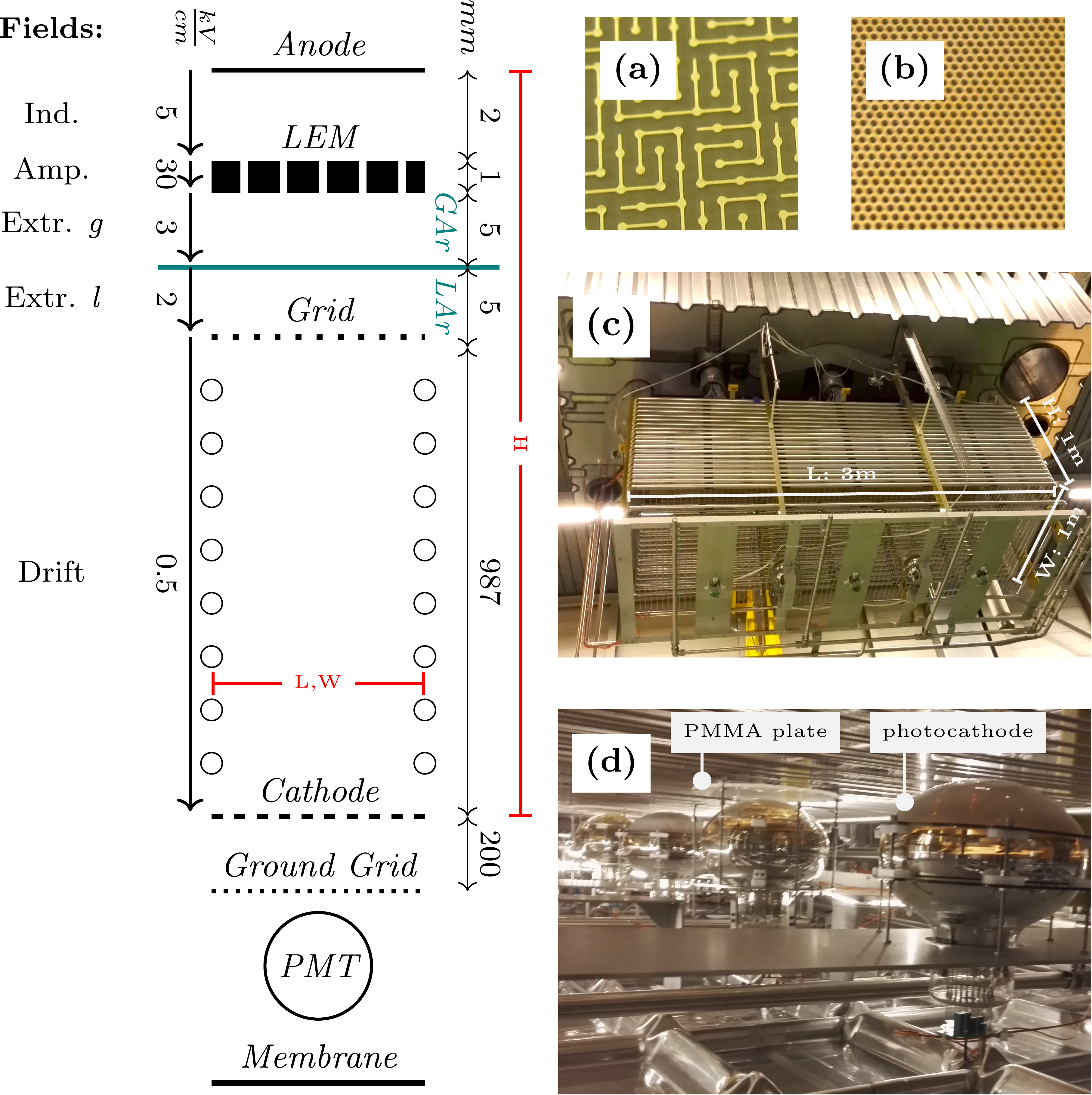}
    \caption{Left: Schematic drawing of the demonstrator elements, with their relative spacing (not to scale) and the nominal values of the Induction, Amplification, Extraction in gas and liquid and Drift fields. On the right, some pictures of the 
    L~$\times$~W~$\times$~H~=~\three detector. (a)~is the segmented anode, (b)~shows the LEM, (c)~is an underneath view of the TPC hung from the top-cap just before its insertion into the cryostat and (d)~displays the PMTs installed underneath the protective ground grid.}
    \label{fig:DP_schema_311_pic}
\end{figure}

The experimental setup used for this paper is shortly reported in this section, however, a complete description can be found in~\cite{311}.
The experimental apparatus is a LArTPC based on the previously introduced dual phase concept. It has an active volume of \SI{3}{\cubic\meter}: a collection surface of \SI{3x1}{\meter} and a drift of \SI{1}{\meter}. 
The whole structure is located in a passively insulated cryostat of about \SI{23}{\cubic\meter} to maintain the argon in liquid state. The main components of the TPC are the field cage including the cathode, the charge readout plane (CRP) and the photon detection system. A schematic view of the TPC and its different components are shown in Fig.~\ref{fig:DP_schema_311_pic}. 

The field cage consists of 20 identical, regularly spaced rings composed of stainless steel tubes. 
Each of the field-forming shaper rings is connected to its neighbors via \SI{100}{\mega\ohm} resistors. The bottom last field shaper forms the cathode composed of a grid of \SI{4}{\milli\meter} diameter stainless steel tubes with a pitch of \SI{20}{\milli\meter}. To achieve a homogeneous electric field -- the drift field -- in the active volume, the voltages of the cathode and the first top field shaper ring can be set independently. 
The required cathode high voltage is generated by a commercial HV power supply and brought from the outside to the cathode through custom-made high voltage feedthrough.

The CRP consists of the extraction grid which completes the drift volume, 12 large electron multipliers (LEMs) to produce the charge avalanche, and the anode to collect the electrons. The extraction grid is composed of \SI{10}{\micro\meter} diameter stainless steel wires with a length of 1 or \SI{3}{\meter}, depending on the X-Y direction. 
The LEMs are installed one centimeter above the extraction grid; the anode is located \SI{2}{\milli\meter} above the LEMs. 
The CRP position is independent from the field cage and remotely adjustable: it is chosen such that the LAr surface lies in the middle of the extraction grid and the LEMs. 
By applying a voltage difference between the extraction grid and the bottom side of the LEMs an electric field -- the extraction field -- is created. It allows to extract the electrons from the liquid to the gaseous phase and guides the electrons towards the LEMs. 
Each LEM consists of a \SI{1}{\milli\meter} thick \SI{50x50}{cm} printed circuit board (PCB) coated on both sides with a thin layer of copper. 
Holes of \SI{500}{\micro\meter} diameter are mechanically drilled in a honeycomb pattern with a pitch of \SI{800}{\micro\meter}. The copper surface around each hole is further removed producing a \SI{40}{\micro\meter}-dielectric rim. By applying a voltage difference between the bottom and the top side of the LEMs a high electric field -- the amplification field -- is produced inside the holes and electrons entering this region undergo charge amplification. 
At the same time, the secondary scintillation light, S2, is generated in the holes. 
The anode plane consists of twelve \SI{50x50}{cm} 4-layer PCB panels. On the bottom layer, orthogonal strips are printed with a \SI{3.125}{\milli\meter} pitch, to provide the two independent views of the event. The pattern of the readout strips is optimised such that the charge is evenly split between both views~\cite{Cantini:2013yba}. 
The electrons drift from the LEM holes to the anode with the final electric field -- the induction field -- and are collected on the strips. Custom-made electronics optimised for the readout of DP LArTPC detectors was used to record the charge signal.

The photon detection system is located below the cathode and an additional grid at ground protecting the photo-sensors from potential discharges. It consists of an array of 5 PMTs distributed along the \SI{3}{\meter} side, separated by \SI{50}{\centi\meter}, installed on the floor of the cryostat. A detailed description of the photon detection system is given in the next section.

\begin{figure}[ht]
\includegraphics[width=15cm]{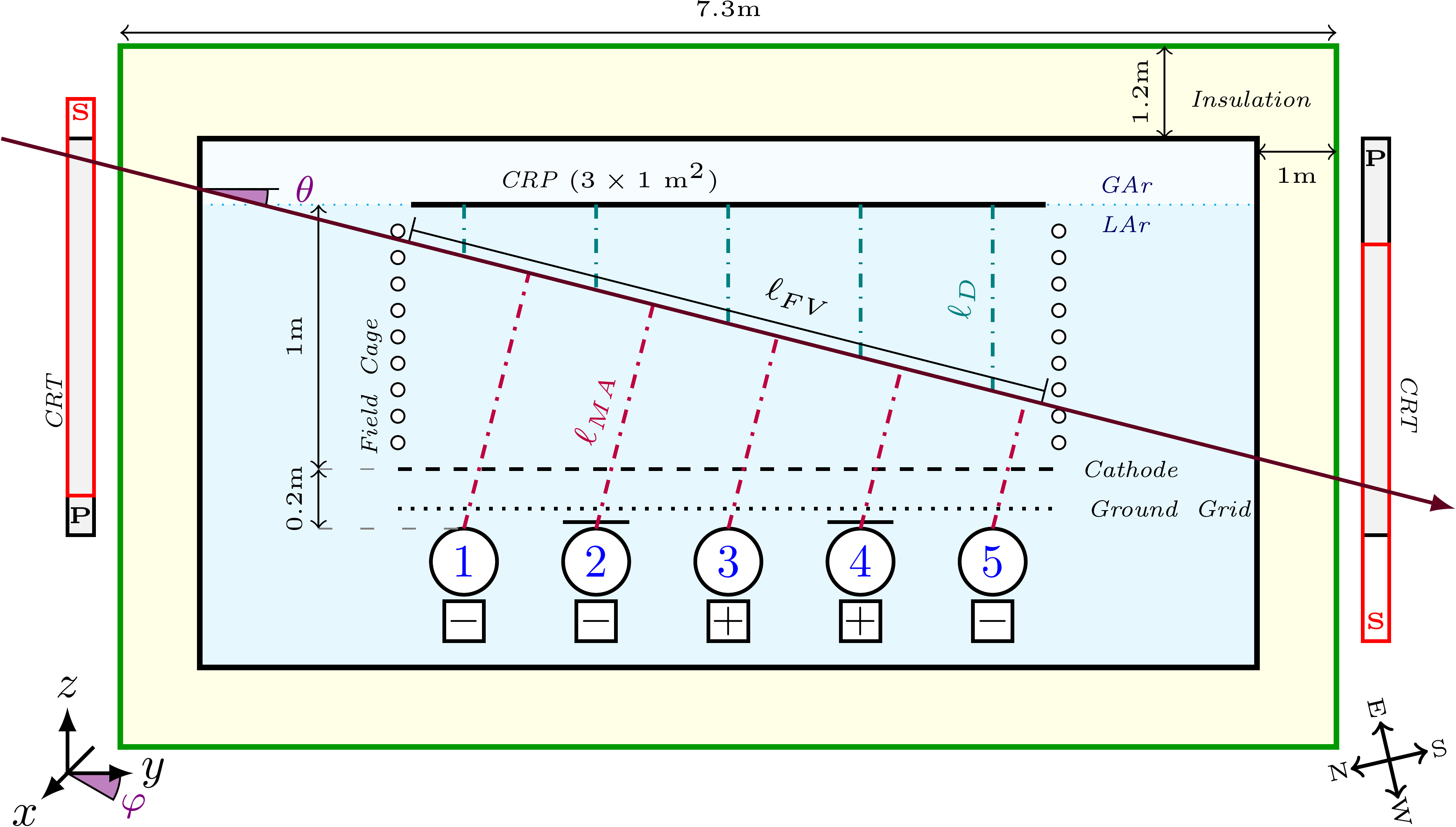}
\caption{\label{fig:311_trigger_tracks}Schematic drawing of the \three demonstrator (not to scale). The different PMT configurations, due to TPB coating methods and electronics polarity, are highlighted. The two CRT positions are noted with a P (parallel) and a S (shifted) label. Relevant track variables used in the analysis are shown.}

\label{fig:311diagram}
\end{figure}

Two movable Cosmic Ray Tagger (CRT) panels were installed on either side of the cryostat external structure along the long side and were used as an external trigger. 
Each of the \SI{1.8x1.9}{\meter} panels was composed of two perpendicular modules of 16 plastic scintillator strips, each having a width of \SI{11}{\centi\meter}, equipped with wavelength shifting optical fibers coupled to two silicon photomultipliers~\cite{Auger}. 
From the fired strips position and time, it is possible to reconstruct the track trajectory and its time of flight, the latter being corrected offline from delays introduced by the cable length and the electronics. During data taking, the position of the CRT panels has been changed once. A schematic view of the \three demonstrator with two CRT configurations and the PMT positions is shown in Fig.~\ref{fig:311_trigger_tracks}. 

Although the performance of the \three demonstrator faced some challenges, it represents an important breakthrough in particle detection with LArTPCs. 
The newly employed membrane cryostat provided a stable cryogenic environment with a flat liquid surface allowing the charge extraction, amplification and collection over an area of \SI{3}{\square\meter}. 
An excellent liquid argon purity was also achieved with a corresponding electron lifetime greater than \SI{4}{\milli\second}. 
While the charge data collection over brief time periods was possible, high voltage instabilities in the operation of the extraction grid prevented a proper study of the detector long-term stability and performance.
The maximum stable amplification field that could be reached in the LEMs was also lower than envisioned for large dual-phase LArTPCs. 
A first look at the data collected at an effective gain of around 3 before complete charging up of the LEMs nevertheless proved the high quality of the dual-phase LArTPC imaging. 
The working principle of the charge readout made of two collection views with strips up to \SI{3}{\meter} length has been demonstrated. Moreover, the extraction of electrons from liquid to gas over a large area has been proven for the first time.

\section{Photon detection system}
\label{sec3}

Five cryogenic Hamamatsu R5912-02MOD PMTs~\cite{protoDUNEPMTs} with an 8-inch diameter borosilicate window were mounted underneath the TPC cathode. The PMTs have a 14-stage dynode chain which provides a nominal gain of 10$^{9}$ at room temperature (RT), required to compensate the gain loss at cryogenic temperatures (CT).  A thin platinum layer was added between the bi-alkali photocathode and the borosilicate glass to preserve the electrical conductivity at CT.

As the borosilicate absorbs light in the VUV range, a thin layer of TPB\footnote{1,1,4,4-Tetraphenyl-1,3-butadiene}~\cite{Bonesini} has been used to convert the \SI{128}{\nano\meter} light to wavelengths in the range of the PMT photocathode sensitivity, the re-emission spectrum of TPB being peaked at \SI{420}{\nano\meter}. To determine the optimum design for future experiments, different wavelength shifter coating techniques and voltage configurations have been considered. Fig.~\ref{fig:311_trigger_tracks} shows the naming convention, the TPB and the electronic base configurations of the 5 PMTs installed in the demonstrator.

Three of the five PMTs were coated with TPB by direct evaporation on the glass window while for the other two, the coating was applied on a transparent \SI{5}{\milli\meter} thick  PMMA\footnote{poly-methyl methacrylate} plate later mounted above the photo-detectors. 
In Fig.~\ref{fig:DP_schema_311_pic}-(d) the two PMT-TPB setups are visible. 

\begin{figure}[h]
\centering
\includegraphics[width=0.95\textwidth]{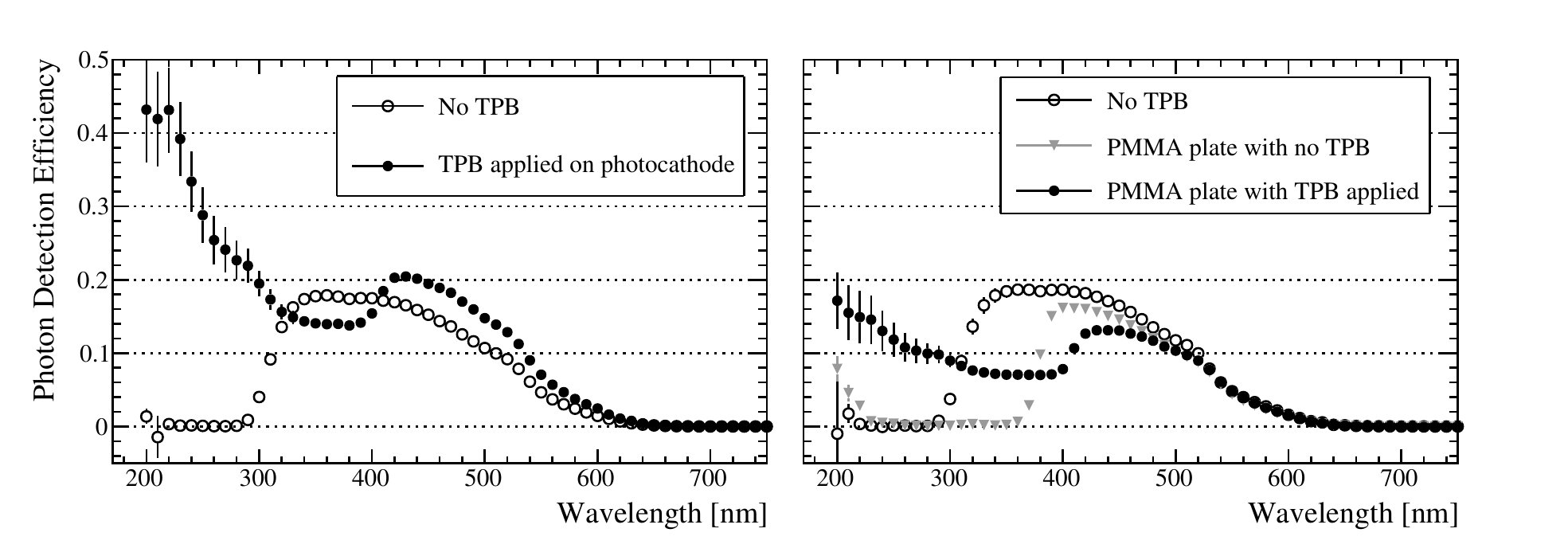}
\caption{Measurement of the PMT photon detection efficiency as a function of the incident wavelength light at RT. Left plot shows the effect of the TPB coated on the PMT photocathode. The effect of placing an acrylic plate above the PMT photocathode with and without TPB is shown in the right plot. The enhancement of the photon detection efficiency at wavelength longer than \SI{400}{\nano\meter} is due light back scattering to the photocathode, as explained in~\cite{Calvo:2016hve}. Given the limitations of the device, these plots should be considered as an illustrative method to compare the TPB coating performance. }
\label{fig:QE}
\end{figure}

Prior to their installation, the photocathode detection efficiency as a function of the wavelength of all PMTs with and without TPB was measured at RT using a dedicated system at CERN. In a dark chamber, one PMT and a calibrated photo-diode were illuminated with a light source in the wavelength range of \SIrange{200}{800}{\nano\meter}. At a given wavelength, the currents of the PMT at the first dynode stage and of the photo-diode was measured. The detection efficiency of the tested PMT could be extracted relatively from the known quantum efficiency of the reference photo-diode. Fig.~\ref{fig:QE} shows the measurements for the two TPB coating methods. In the absence of wavelength shifter, the photocathode detection efficiency for \SI{420}{\nano\meter} photons at RT is at $\sim$20\%, which is in agreement with the specification provided by the manufacturer. As clearly seen in Fig.~\ref{fig:QE}, once the TPB was applied, the PMTs became sensitive to the UV light. However, in this wavelength range (below \SI{300}{\nano\meter}), the measured values were subject to large fluctuations with time with no clear reasons found. Given this limitation, the measured values presented in this figure were not used neither in the simulations nor in the analysis.

Nevertheless, as seen in Fig.~\ref{fig:QE}, the performance are clearly better when the TPB is directly deposited on the PMT. This is due to several aspects: a larger acceptance, possible loss of photons due to total reflections at different optical interfaces and acrylic light absorption. 
However, in terms of risks and practicality, the plates present enormous advantages at the coating, handling, storing and installation stages. 

Two PMT base designs have been tested, respectively named positive and negative bases due to the sign of the bias voltage applied. In the negative base (NB), a negative voltage is applied at the photocathode and the anode is grounded. Two cables are required: one for the HV and one for the signal readout. 
The positive base (PB) operates with a positive bias voltage applied to the anode while the photocathode is grounded. The dark current induced by spurious pulses from current leakage through the glass could be reduced in this setup.
Moreover, the PB needs only one coaxial cable for carrying both the HV and the PMT signal. The two signals are split with a decoupling circuit outside of the cryostat at RT. It is composed of a RC low-pass filter to remove noise from the power supply and by a capacitor\footnote{\SI{200}{\nano\farad} for PMT 3, \SI{300}{\nano\farad} for PMT 4} that decouples the PMT signal from the DC power supply.

The light readout signals were acquired using commercial electronics.
The analog signals from each PMT were digitised up to \SI{1}{\milli\second} window. 
The digitisation was performed with a resolution of \SI{12}{bit} sampled at \SI{250}{\mega\hertz} using a CAEN~v1720 board. 
The ADC has a total dynamic range of \SI{2}{\volt}, limiting the set PMT gain during operation.
The board was read out via an optical link to a PC equipped with a CAEN~A2818 PCI CARD. The software for the display and acquisition was based on the MIDAS framework~\cite{midas} and ran on the same PC. 
The board also allowed to program a simple majority coincidence trigger.

\begin{figure}[ht]
\centering
\includegraphics[width=0.9\textwidth]{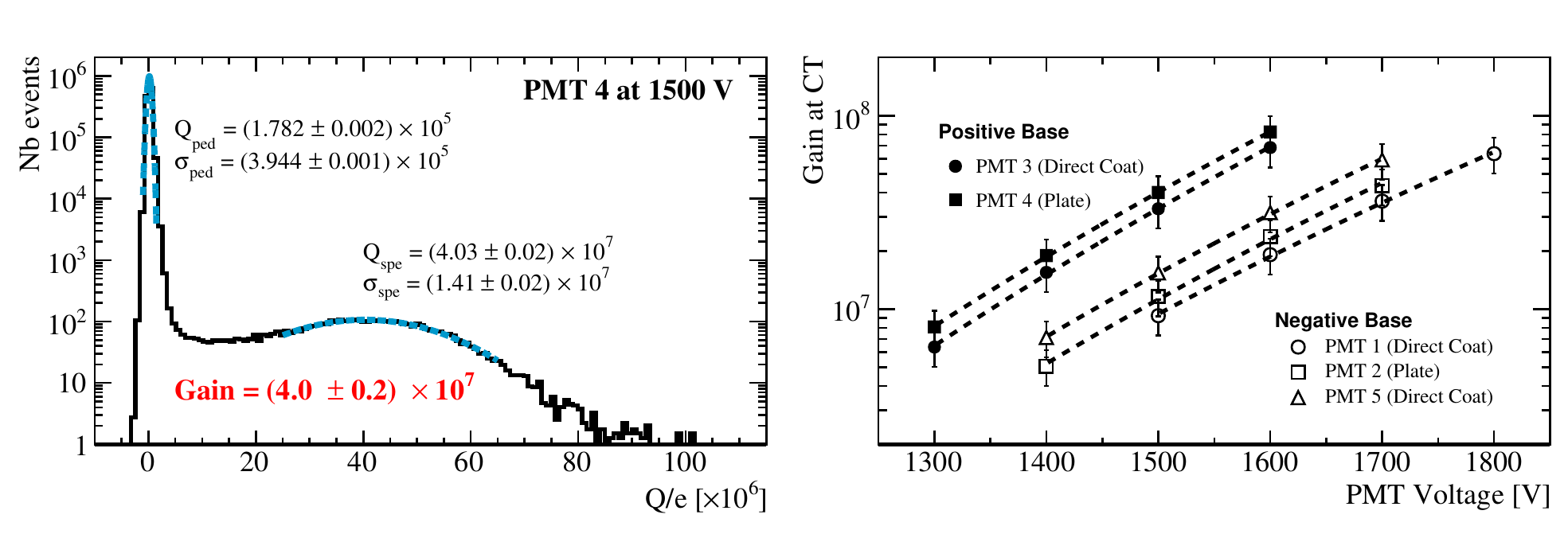}
\caption{Left: Integrated charge spectrum for one PMT of the single PE analysis and gain computation. Right: PMT gain measured in liquid argon for different operating voltages. The calibration could not be performed at lower voltages as the single PE signal was indistinguishable from the pedestal fluctuation. The fitted power law is superimposed.}
\label{fig:gain}
\end{figure}

Once the PMTs were immersed in LAr, a voltage scan has been performed to obtain the gain calibration curve for every PMT using a \SI{100}{\hertz} pulse generator as a trigger. 
Every fluctuation from the baseline along the waveform is integrated to get the corresponding charge.  
The obtained spectrum consists of two Gaussian distributions, one centered at zero corresponding to the pedestal while the other represents the PMT response to a single photo-electron (PE). The PMT gain is calculated as the distance between the means of the two Gaussian distributions, expressed in units of the electron charge.
The dependence of the gain ($G$) on the voltage ($V$) applied to the PMT is fitted to a power law $G = AV^{B}$, where $A$ and $B$ are constants depending on the number, structure, and material of the dynodes. An example of the fitted charge spectrum and the PMT gain calibration curves are shown in Fig.~\ref{fig:gain}. 
The PMTs were initially operated at a gain of 10$^{7}$, and later  decreased to 10$^{6}$ in order to reduce ADC saturated events.

\begin{figure}[ht]
\begin{center}
\includegraphics[width=0.45\textwidth]{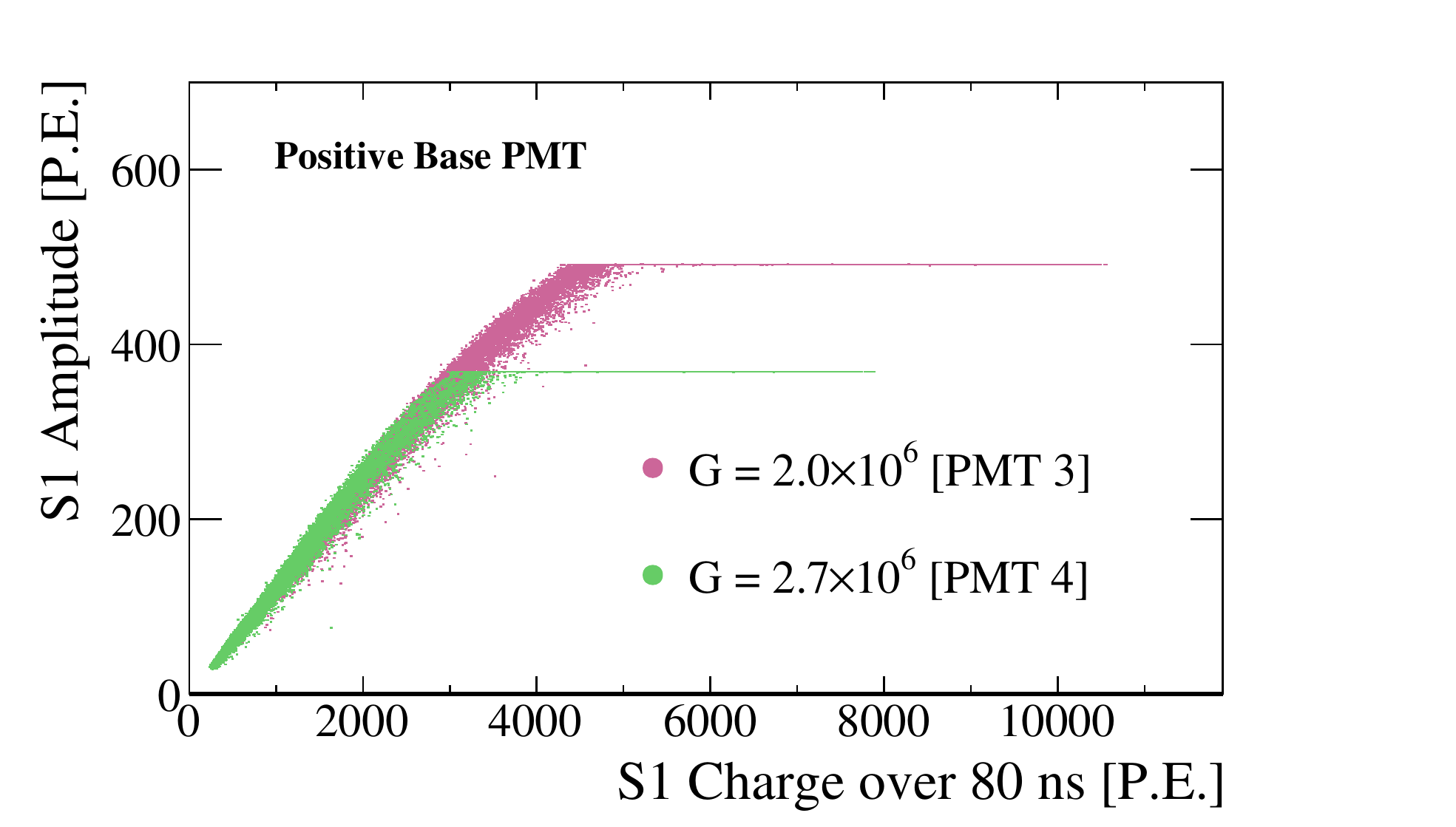}
\includegraphics[width=0.45\textwidth]{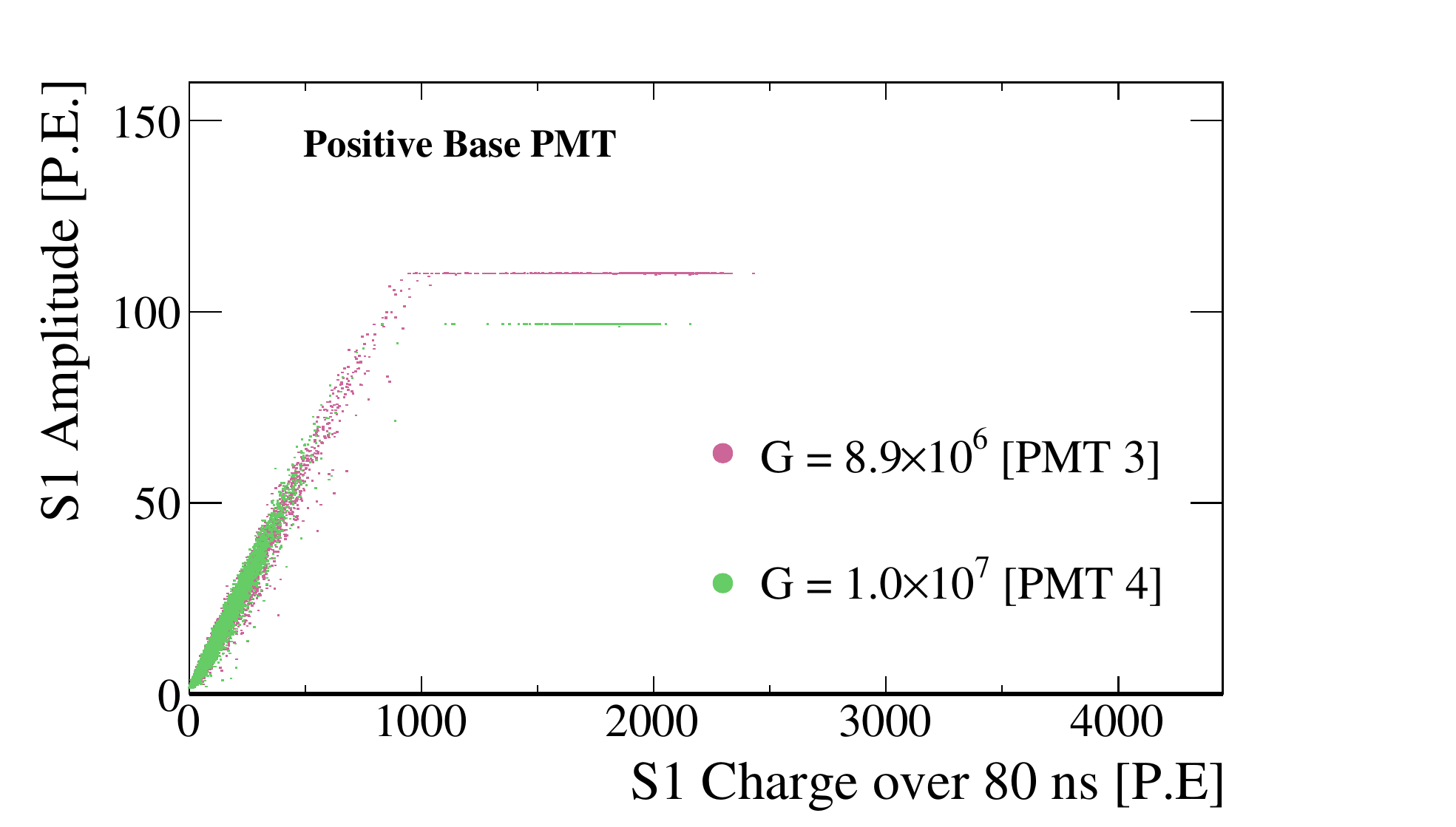}\\
\includegraphics[width=0.45\textwidth]{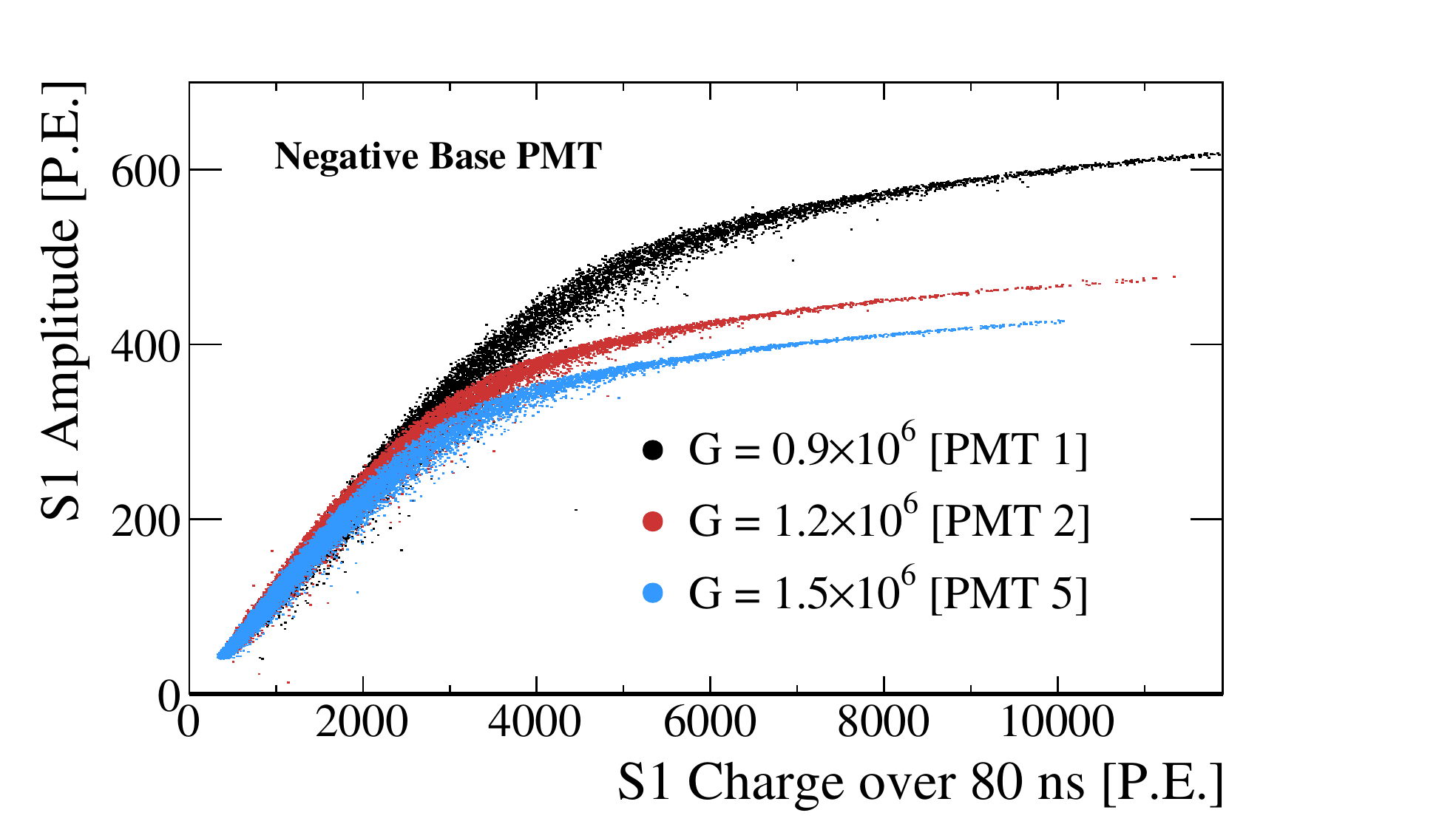}
\includegraphics[width=0.45\textwidth]{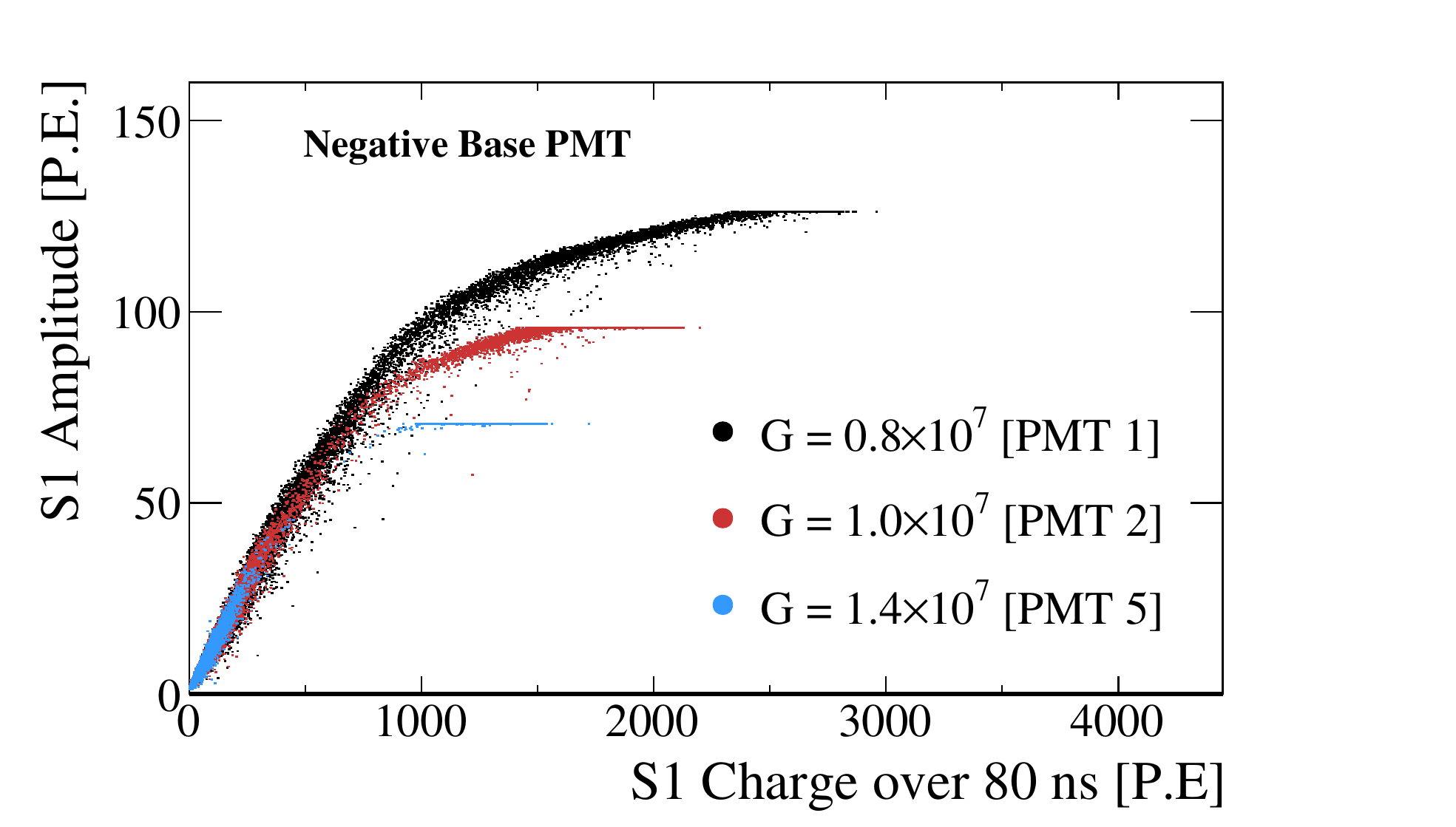}
\caption{The maximum amplitude of the S1 signal versus S1 fast pulse integrated charge over \SI{80}{\nano\second} in PE is shown for the PB (top) and NB (bottom) PMTs, for gains around 10$^{6}$ (left) and 10$^{7}$ (right).}
\label{fig:satur}
\end{center}
\end{figure}

The PMT response at CT has been studied using the collected S1 signals. In particular, the linearity between the sampled charge at the waveform maximum amplitude and integrated over \SI{80}{\nano\second} is shown in Fig.~\ref{fig:satur}. 
For amplitudes larger than $\sim$\SI{300}{PE} ($\sim$\SI{75}{PE}) at a gain of 10$^6$ (10$^7$), the linear relationship is lost for NB PMTs before the ADC saturation. On the contrary, the PB PMT linearity is preserved almost in the whole allowed ADC range at both gain settings.
This effect has been investigated with similar PMTs and electronic bases at RT and CT in~\cite{protoDUNEPMTs}. In these studies, using a \SI{40}{\nano\second} pulsed LED, the linearity was lost for a charge higher than \SI{200}{PE} at a gain of 10$^7$ in CT. It was also shown that the shorter the pulse, the earlier the linearity is lost. These results are in agreement with what is observed in the \three data, and points at an intrinsic effect of the light collection system and discard physics processes affecting the shape of the scintillation light profile.

\begin{figure}[ht]
\begin{center}
\includegraphics[width=0.45\textwidth]{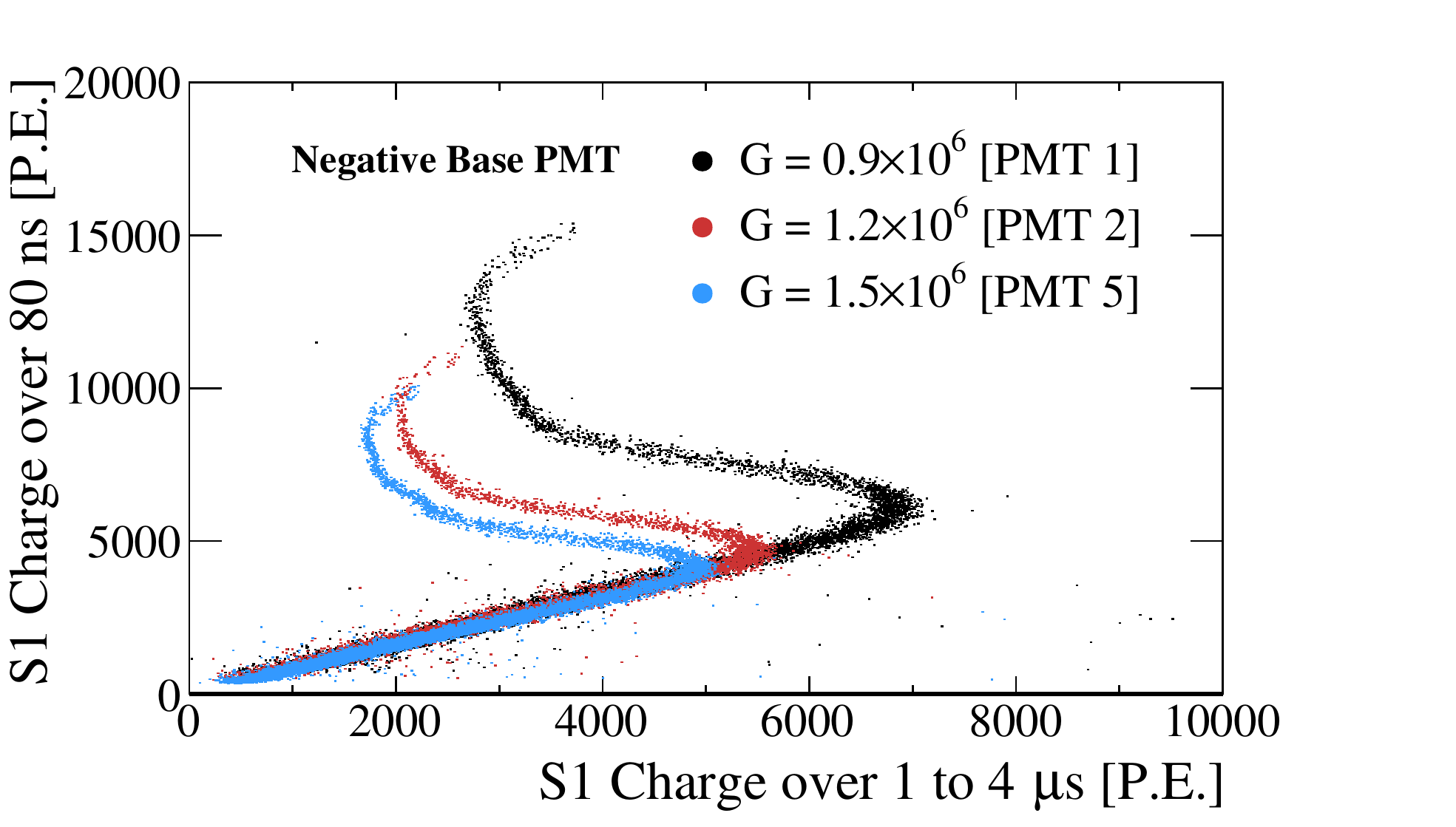}
\includegraphics[width=0.45\textwidth]{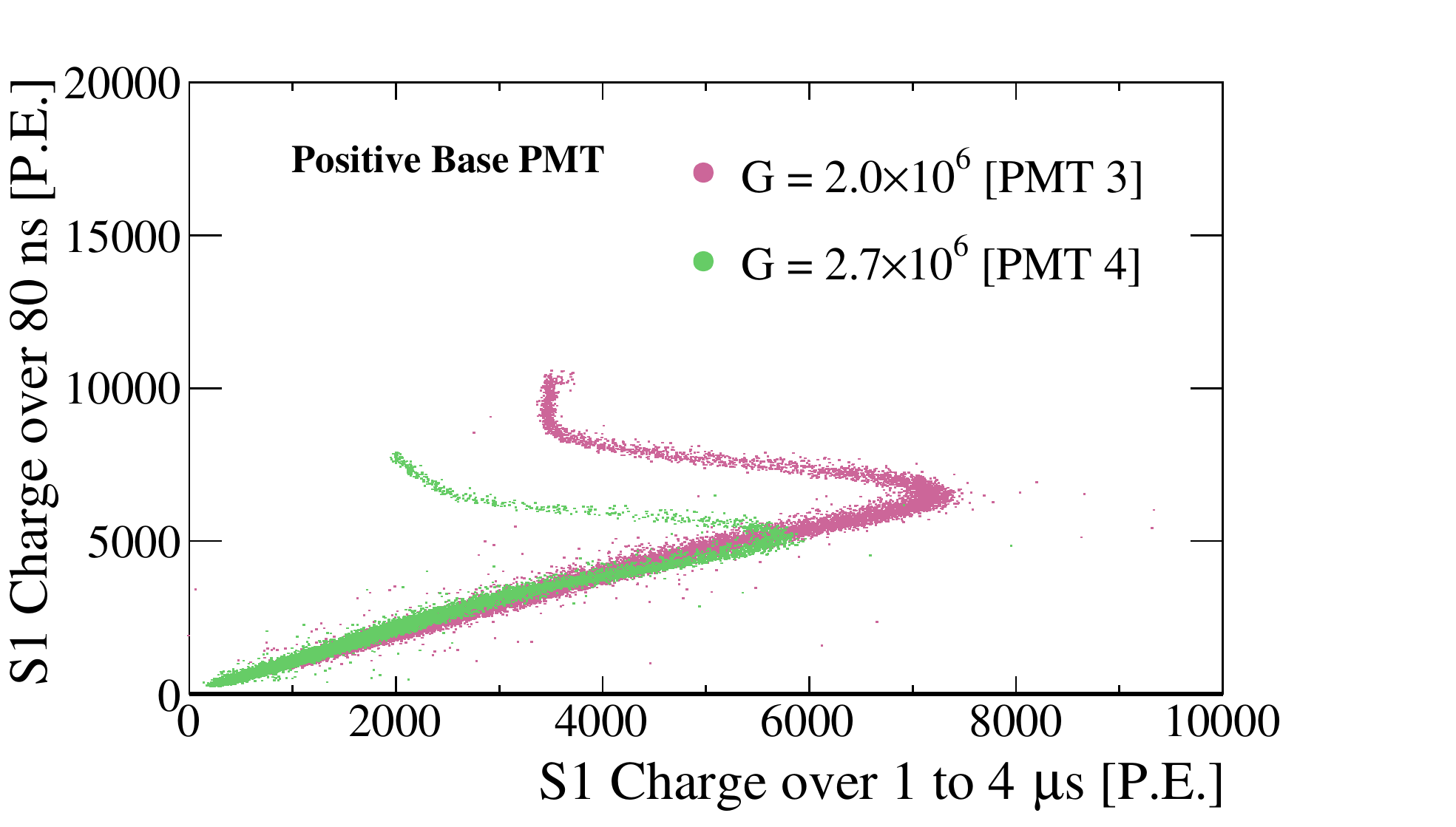}
\caption{S1 fast pulse integrated charge versus the integrated charge corresponding to the slow component of the waveform, both in PE. The effect is visible in NB PMTs (left) and PB ones (right) for gains around 10$^6$.}
\label{fig:comet}
\end{center}
\end{figure}

\begin{figure}[ht]
\centering
\includegraphics[width=0.9\textwidth]{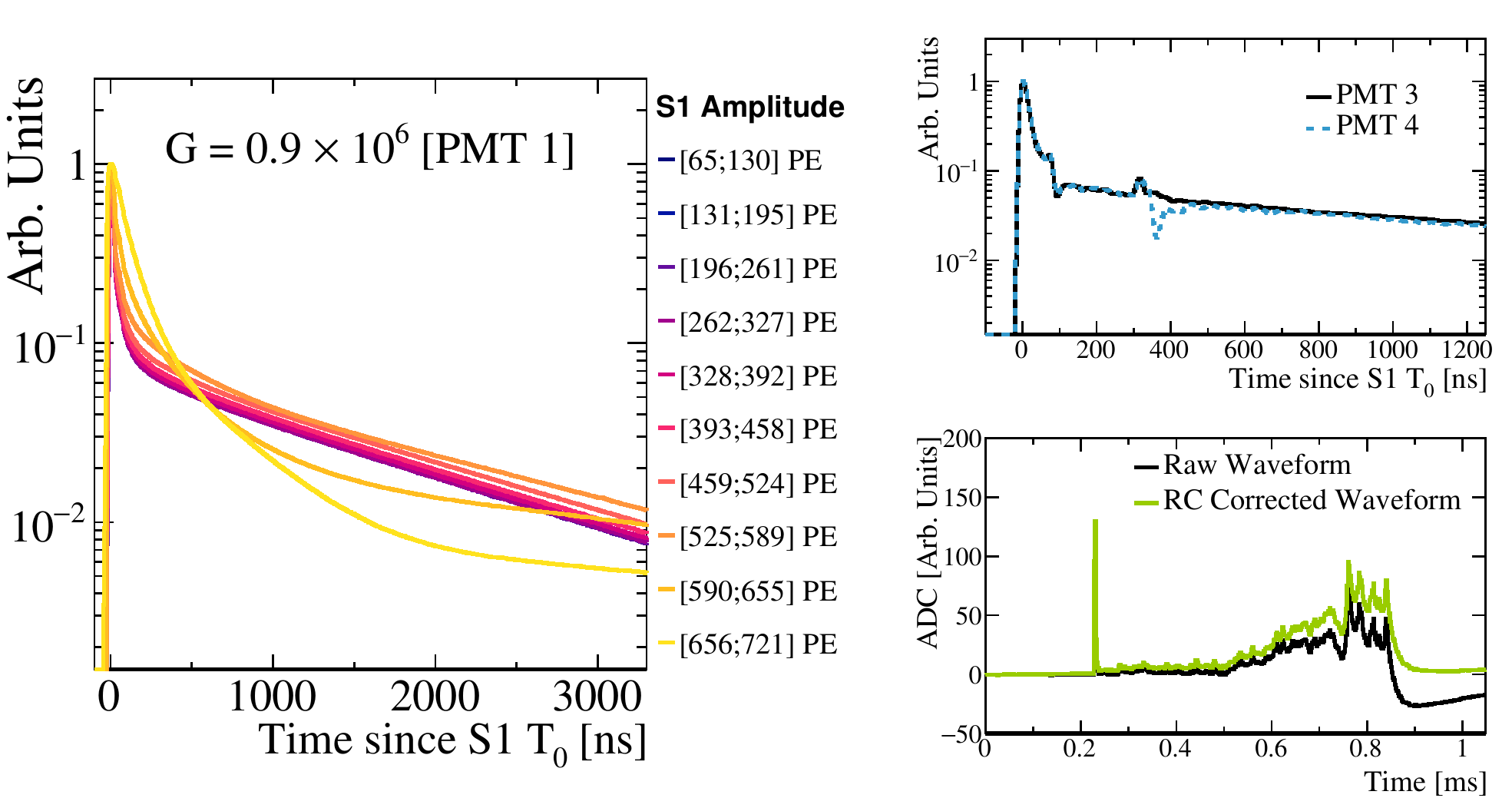}
\caption{Left: Averaged normalized waveforms for one PMT in slices of S1 maximum amplitude (in range of \SI{250}{ADC} corresponding to \SI{66}{PE}). For reference, in the analysis for this PMT at this gain, any event with an S1 amplitude higher than 365 PE is discarded. Right top: Averaged normalized waveforms for PMTs with a positive base around the S1 peak. The signal reflection $\sim$\SI{100}{\nano\second} and $\sim$\SI{350}{\nano\second} after the peak is clearly visible. Right bottom: One long waveform with the S1 and S2 signals collected by a PMT with a positive base. The overshooting of the S2 signal is seen in black and corrected offline in green.}
\label{fig:waveforms}
\end{figure}

Similarly, the linearity between the charge integrated over \SI{80}{\nano\second} (the fast part) and over \SIrange{1}{4}{\micro\second} (the slow part) after the waveform maximum amplitude is shown in Fig.~\ref{fig:comet}. 
The scintillation light profile is distorted for very large signals: while the fast part increases, the slow part decreases up to a factor 2.5, see Fig.~\ref{fig:waveforms}-left.
The more probable hypothesis is a saturation of the anode of the PMT, unable to provide the output signal for high frequency pulses.  The S1 slow component consists in trains of large pulses over a relative long period ($\sim$\SI{2.5}{\micro\second}). It was proven in~\cite{protoDUNEPMTs} that the PMT saturates with high light frequency, even for small amplitudes at few PEs level.  Those pulses produce a high average output current, so, once the capacitors on the last dynode are discharged after the first pulses, they cannot recover until the end of the pulse train, loosing part of the collected charge.

Both effects (PMT linearity and saturation) seem to be either linked or induced by events in the similar energy range. For the analyses presented in this paper, only events having an S1 highest amplitude lower than a value defined for each PMT at each gain setting are kept. For data taken at a gain of $10^6$, about 15\% of the events are rejected.

For PB PMTs, a mismatch impedance in the cryostat feedthrough generated signal reflections clearly visible after the S1 peak in Fig.~\ref{fig:waveforms}-top right. Furthermore, large S2 pulses were observed to cause overshooting in the waveforms of PB PMTs. This occurs when the amount of charge collected in the PMT anode exceeds the discharging rate of the combined PMT and readout circuit (1/$RC$ constant), effectively shifting the waveform baseline during the pulse. Waveforms were processed offline to correct for this effect, as shown in Fig.~\ref{fig:waveforms}-bottom right. Since the charging and discharging rate, $RC_{c}$ and $RC_{d}$, vary depending on the profile of the S2 signal (i.e. duty cycle), a best-fit effective $RC$ constant, $RC_{eff} = RC_{c} = RC_{d}$ was calculated for each waveform and used to correct the baseline fluctuation due to charging and discharging bin-by-bin. The correction procedure was tested and qualified using a stand-alone test setup in the laboratory.

\begin{figure}[ht]
\begin{center}
\includegraphics[width=0.8\textwidth]{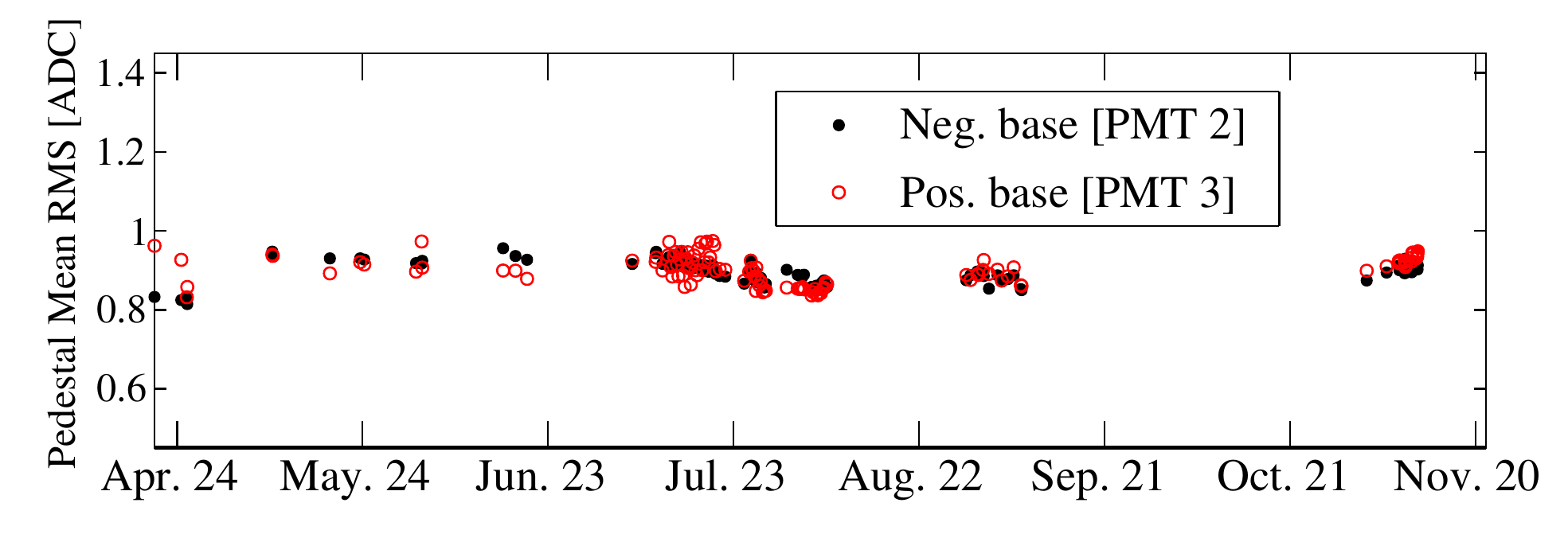}
\caption{Pedestal Mean RMS of each run for one PB (open red) and one NB PMT (full black).}
\label{fig:stab}
\end{center}
\end{figure}

Finally, concerning the noise, the photon detection system has shown very stable performance during data taking.  In Fig.~\ref{fig:stab}, the pedestal RMS is shown for one positive and one negative PMT base as a function of time, for several months of operation. The pedestal RMS is measured to be around \SI{1}{ADC} count for both bases. 

\section{Data description}
\label{sec4}

The \three demonstrator took charge data from June to November 2017.  Starting April 2017, using the trigger given by the CRT system, scintillation light data were recorded during the purging, cooling down and LAr filling stages prior to the commissioning of the whole detector. 

\begin{figure}[h]
    \centering
    \includegraphics[width=0.45\textwidth]{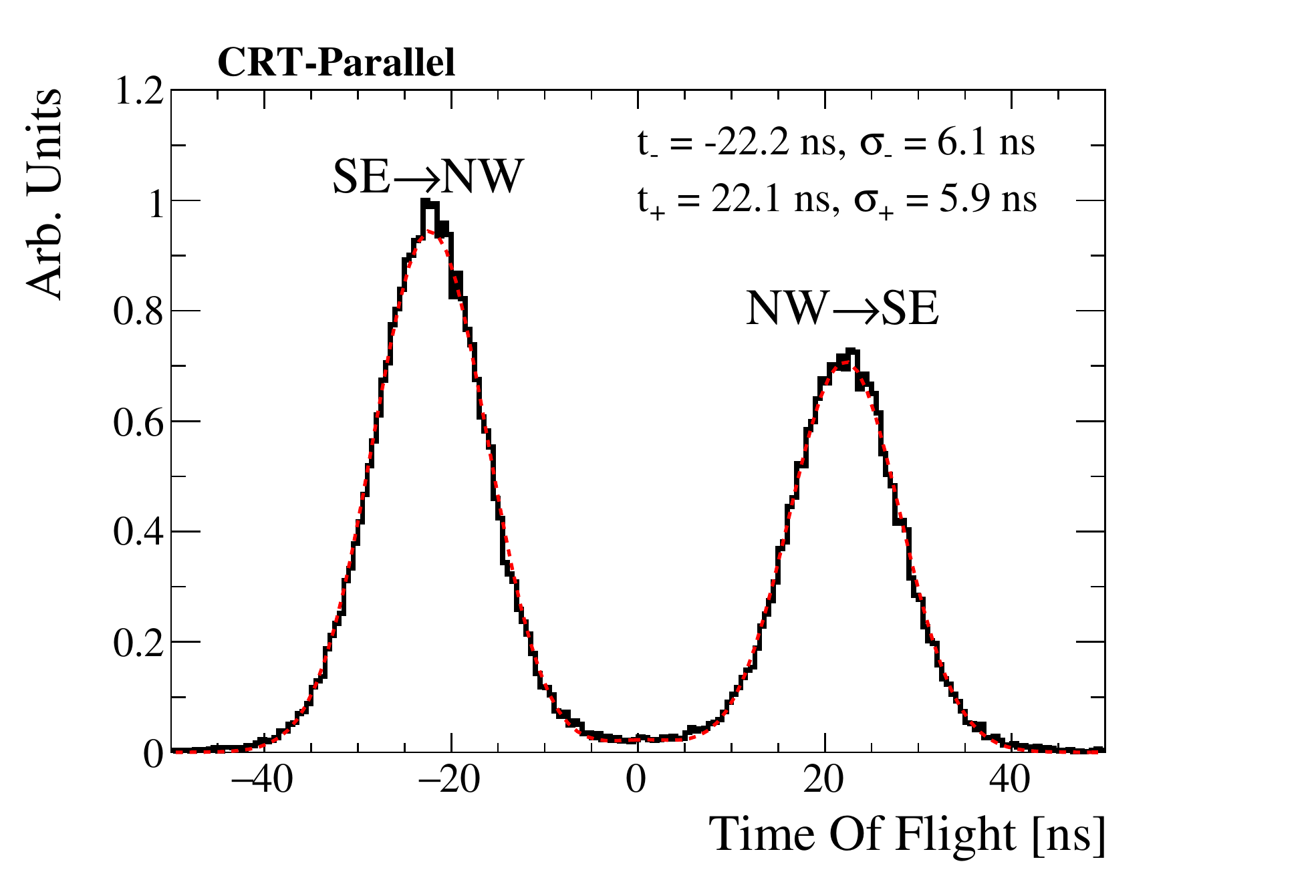}
    \includegraphics[width=0.45\textwidth]{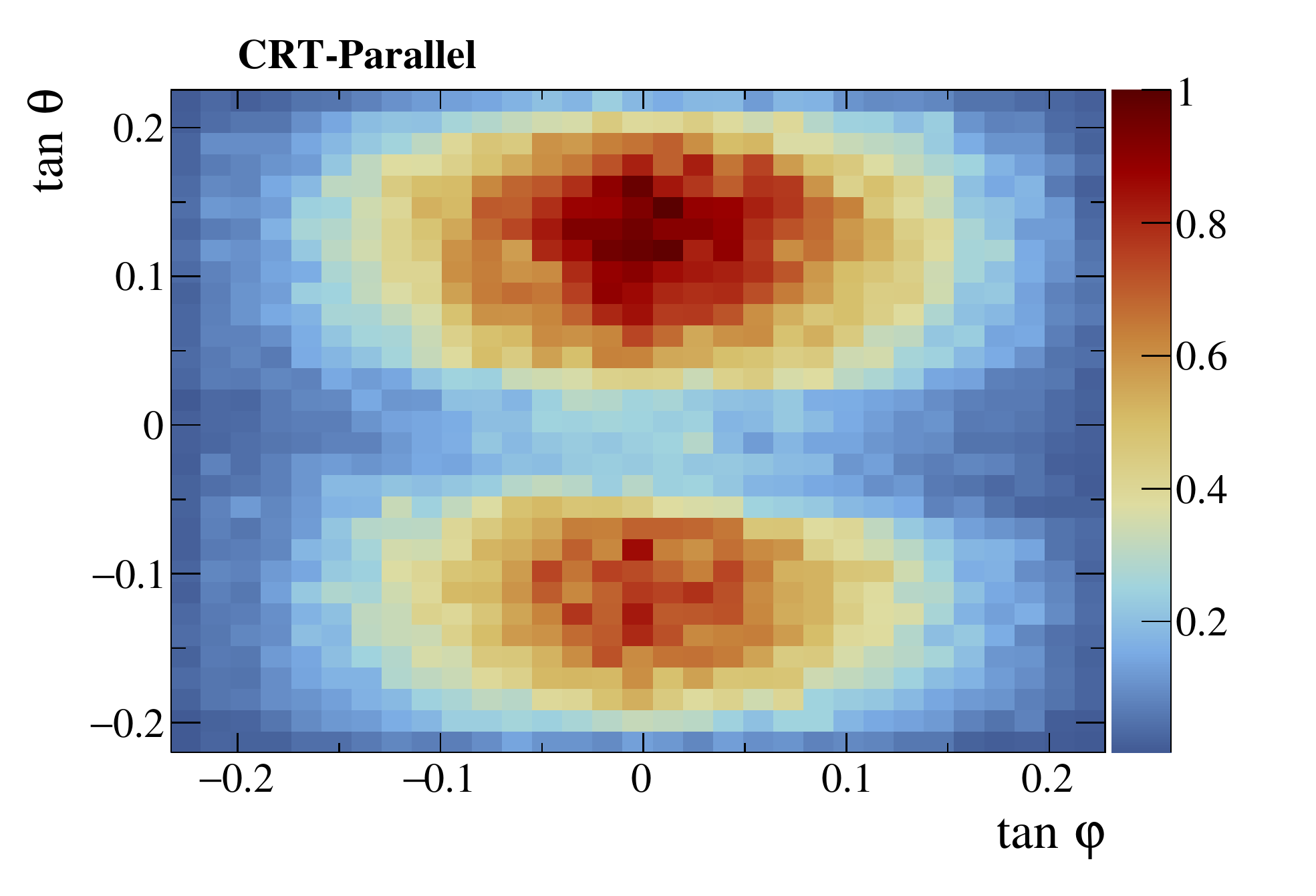}\\
        \includegraphics[width=0.45\textwidth]{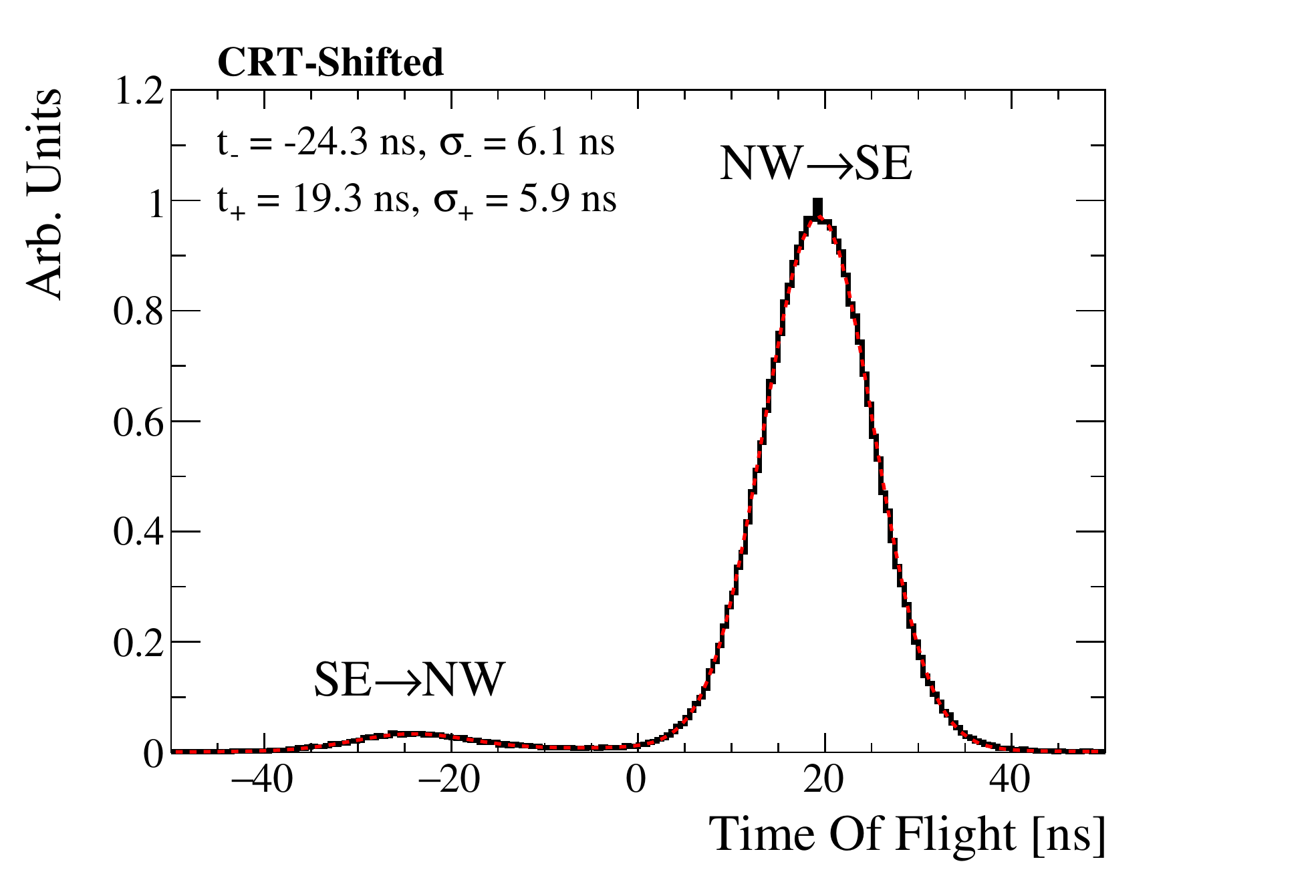}
    \includegraphics[width=0.45\textwidth]{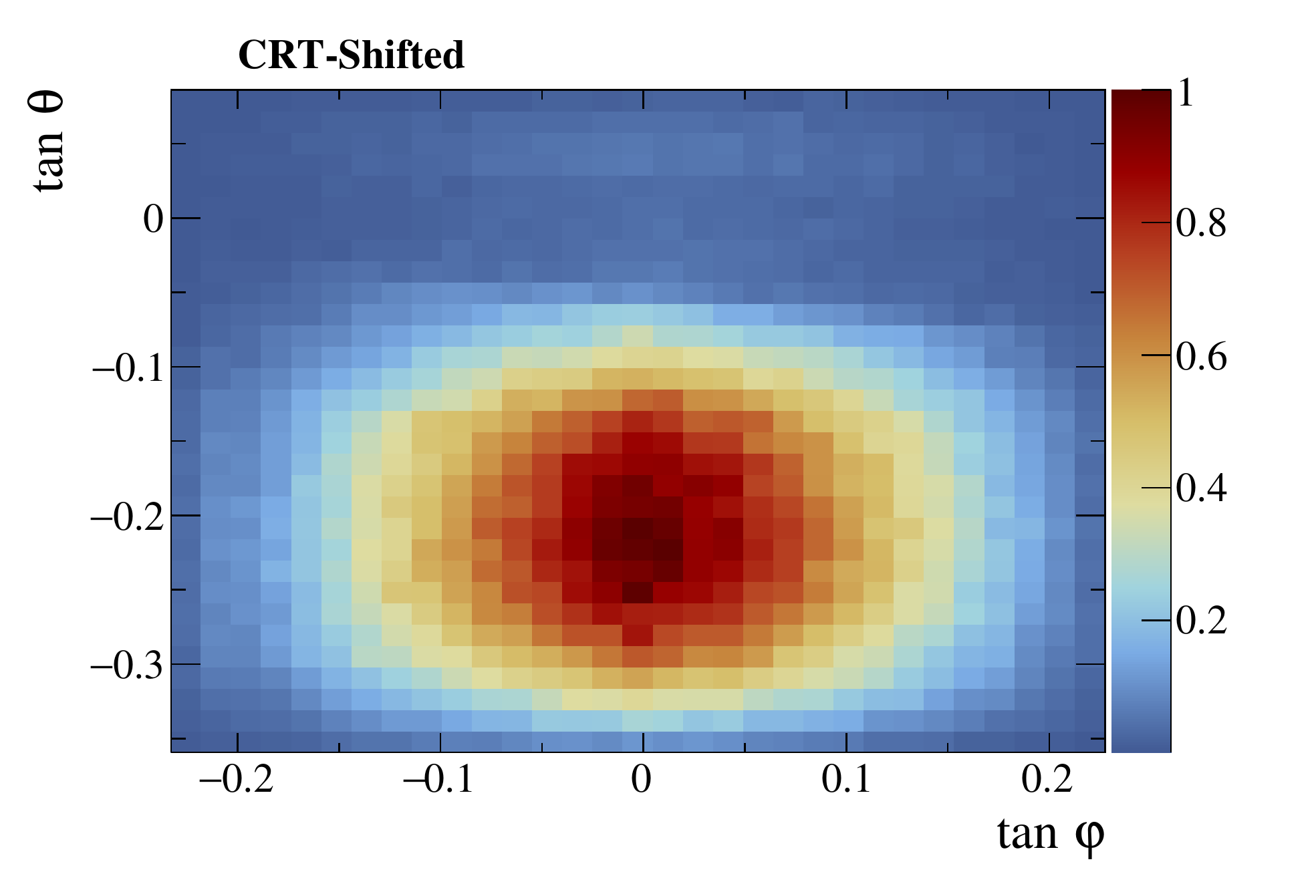}

    \caption{Time of flight (left) and ($\theta$-$\varphi$) phase space (right) of muon-like tracks triggered by the CRT system in parallel (top) and shifted position (bottom). The effect of the Jura mountains (located on the north-west) can be seen on all distributions. }
    \label{fig:crt-trigger}

\end{figure}

Two CRT positions were used for the analysis.
The first positioning of the CRTs was determined to record tracks parallel to the CRP plane, a topology similar to the beam events that will be taken in DUNE. As the flux of cosmic muons follows a $\sin^2\theta$ law, the rate of muon-like tracks triggered in this configuration, later referred to as ``CRT-parallel'', was of the order of \SI{0.3}{\hertz}.  In an attempt to increase the trigger rate, the CRT panels were moved vertically to an asymmetrical configuration, later called ``CRT-shifted'' (+\SI{15}{\centi\meter}; \SI{-90}{\centi\meter}).  Even though the trigger rate remained roughly as low as previously, the CRT-shifted configuration allowed us to record numerous muon-like tracks crossing the entire drift volume from the anode to the cathode.  This topology is handy for many analyses, as no T$_0$ correction is needed. The time of flight and ($\theta$-$\varphi$) phase space of muon-like events triggered by the CRTs in both positions is shown in Fig~\ref{fig:crt-trigger}.

\begin{figure}[h]
    \centering
    \includegraphics[width=0.8\textwidth]{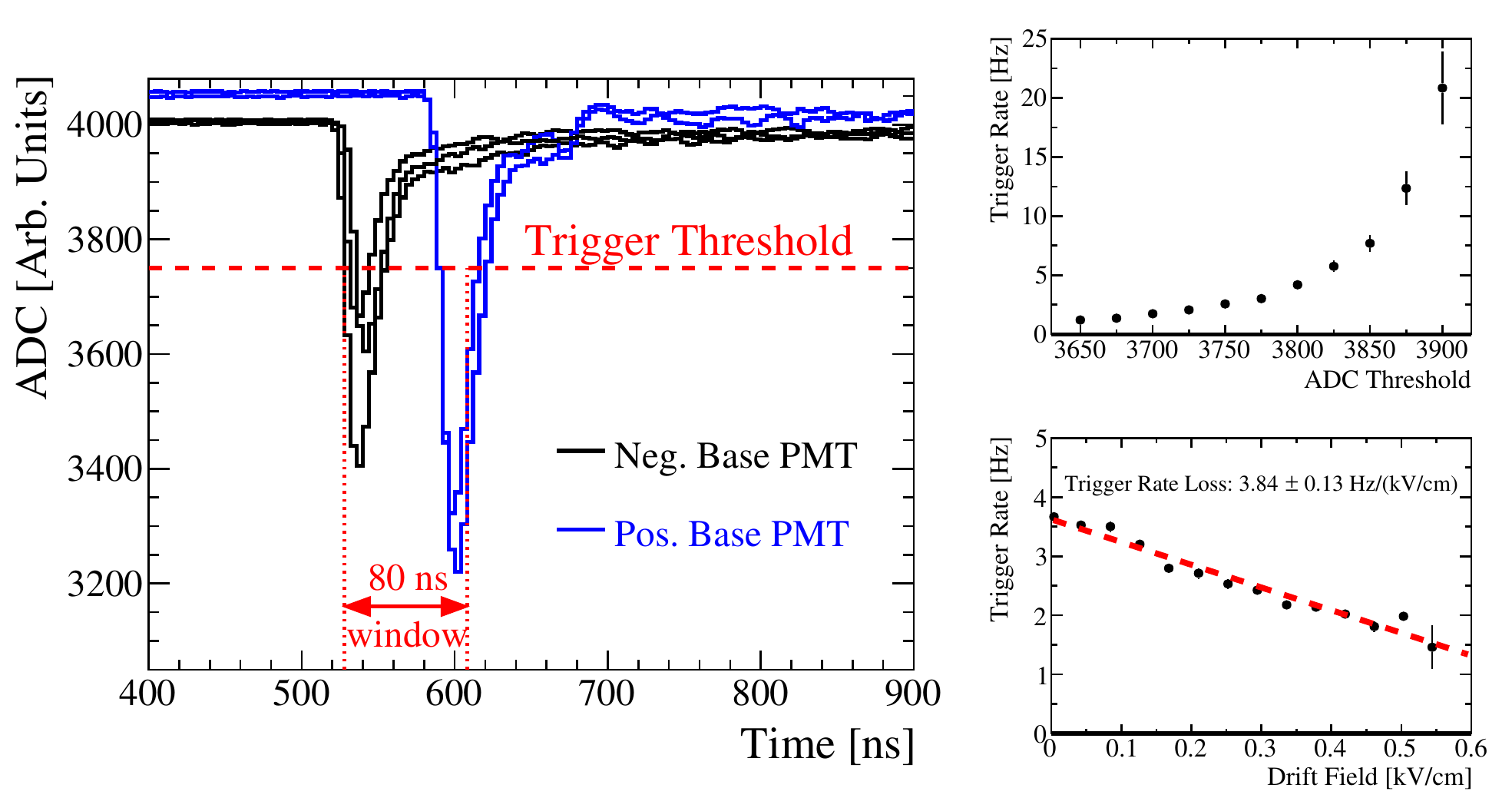}
    \caption{Left: PMT self trigger configuration illustrated with one raw event. Top-right: Trigger rate as a function of ADC threshold set to all PMTs taken at a null drift field. Bottom-right: Trigger rate as a function of drift field with a constant threshold set at \SI{3750}{ADC} on all PMTs.}
    \label{fig:pmt-trigger}

\end{figure}

The other trigger condition tested in the demonstrator was done by the time coincidence (within a \SI{80}{\nano\second} window) of the 5~PMTs themselves.
Standard operation conditions corresponded to an ADC threshold set to the PMTs adjusted for a trigger rate at a drift field of \SI{500}{\volt/\centi\meter} below \SI{3}{\hertz}. 
In Fig.~\ref{fig:pmt-trigger}, the trigger configuration and a study of the trigger rate as a function of the ADC cut is shown.  In PMT self trigger mode, not only single tracks were collected, showers and multi-tracks events were also recorded.  

The \three demonstrator allowed the characterization of both S1 and S2 signals. 
A large amount of data were collected with a \SI{4}{\micro\second} acquisition window to study the primary scintillation light. In this configuration, a pre-trigger of \SI{0.5}{\micro\second} was set for an event by event determination of the pedestal. 
In presence of extraction and amplification fields, the data was collected with a \SI{1}{\milli\second} acquisition window (the pre-trigger was then set to \SI{0.2}{\milli\second}) to fully characterize the S2 signal. The acquisition window is then longer than the time it takes for an electron to drift over \SI{1}{\meter} at the nominal drift field of \SI{500}{\volt/\centi\meter} ($\sim$\SI{625}{\micro\second}).

\begin{figure}[ht]
    \centering
    \includegraphics[width=\textwidth]{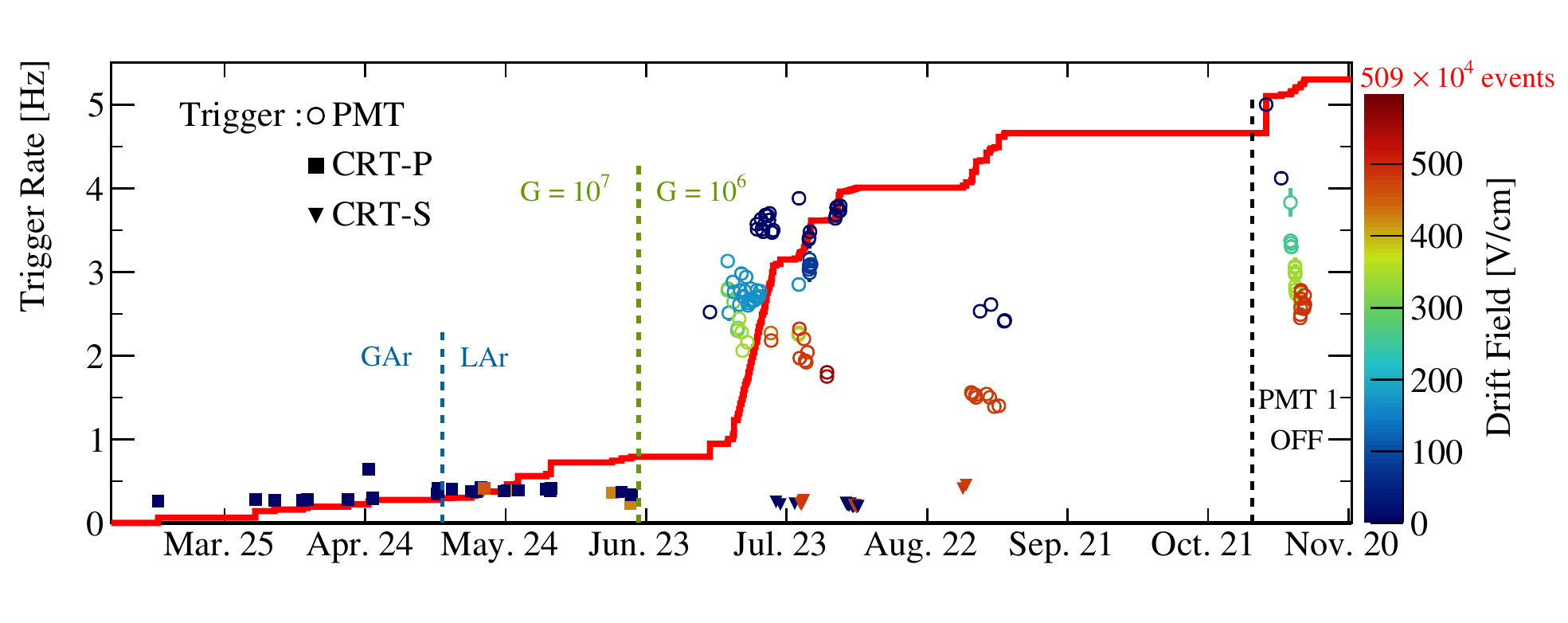}
    \caption{Summary of the scintillation light data taken. One point corresponds to one run in a given HV condition of the detector. The three trigger conditions are noted with different labels. The color of the points represents the drift field value of the run from which its effect on the trigger rate can be seen. The red line represents the cumulative number of collected events.}
    \label{fig:Nevents-rate}
\end{figure}

The \three demonstrator collected data in various configurations of drift, extraction and amplification fields.
In Fig.~\ref{fig:Nevents-rate}, the trigger rate of all physics runs is shown as a function of time, trigger condition and drift field. The cumulative number of collected events is also presented. 
In PMT self trigger mode, due to the recombination of the electrons with argon ions, the trigger rate decreases with increasing drift field at a fixed ADC threshold. 
In order to study this effect, a dedicated set of short runs was taken in a time span of 2~hours, in drift field steps of around \SI{50}{\volt/\centi\meter} from \SIrange{0}{560}{\volt/\centi\meter}, shown in Fig.~\ref{fig:pmt-trigger}-right.
In the later runs taken in November, due to a faulty cable, the PMT~1 was turned off. The PMT self-trigger mode was then set by the time coincidence of the remaining four PMTs, increasing the trigger rate.

\section{Light simulation}
\label{sec5}

\begin{figure}
    \centering
    \includegraphics[width=0.32\textwidth]{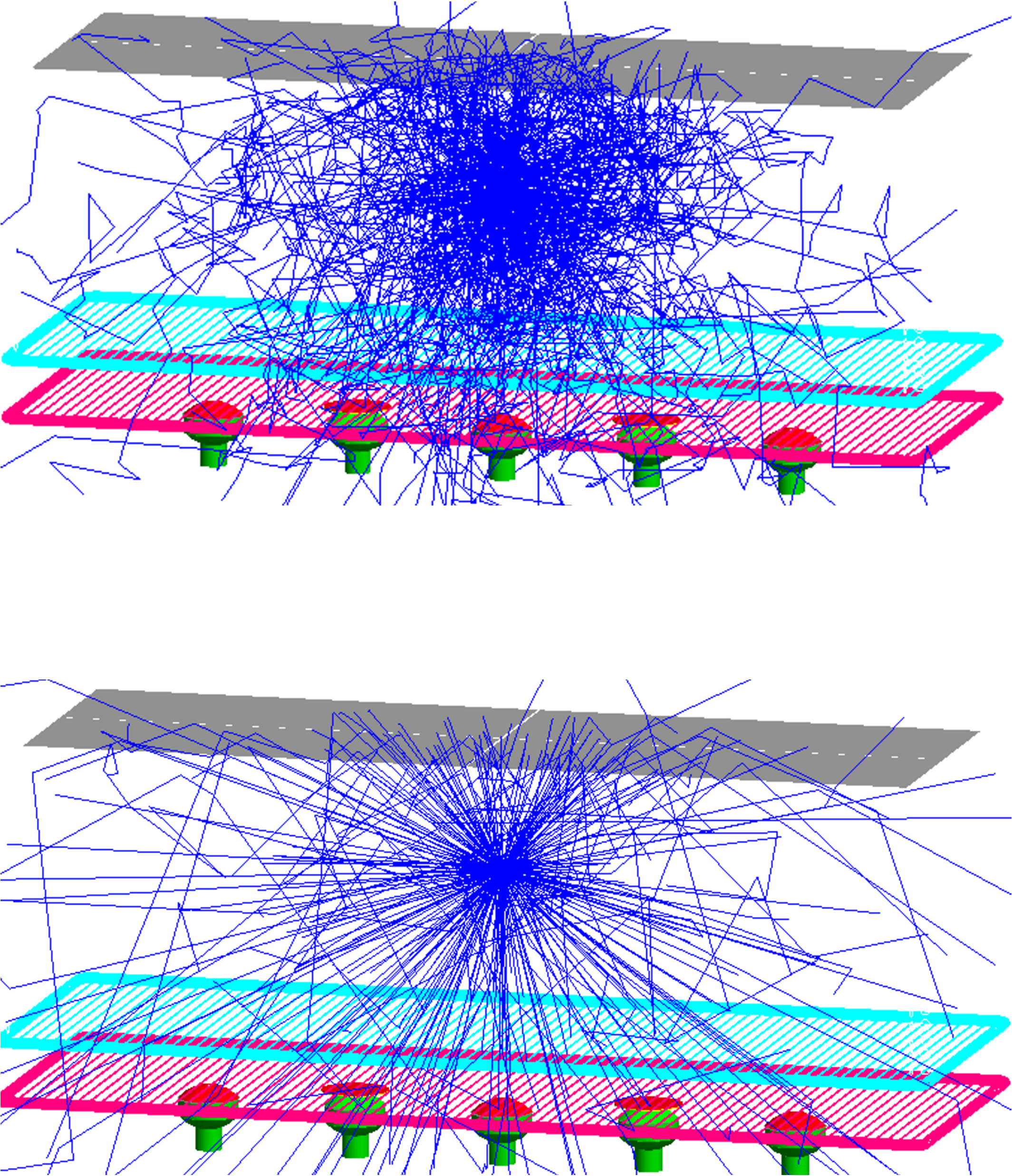}
    \includegraphics[width=0.5\textwidth]{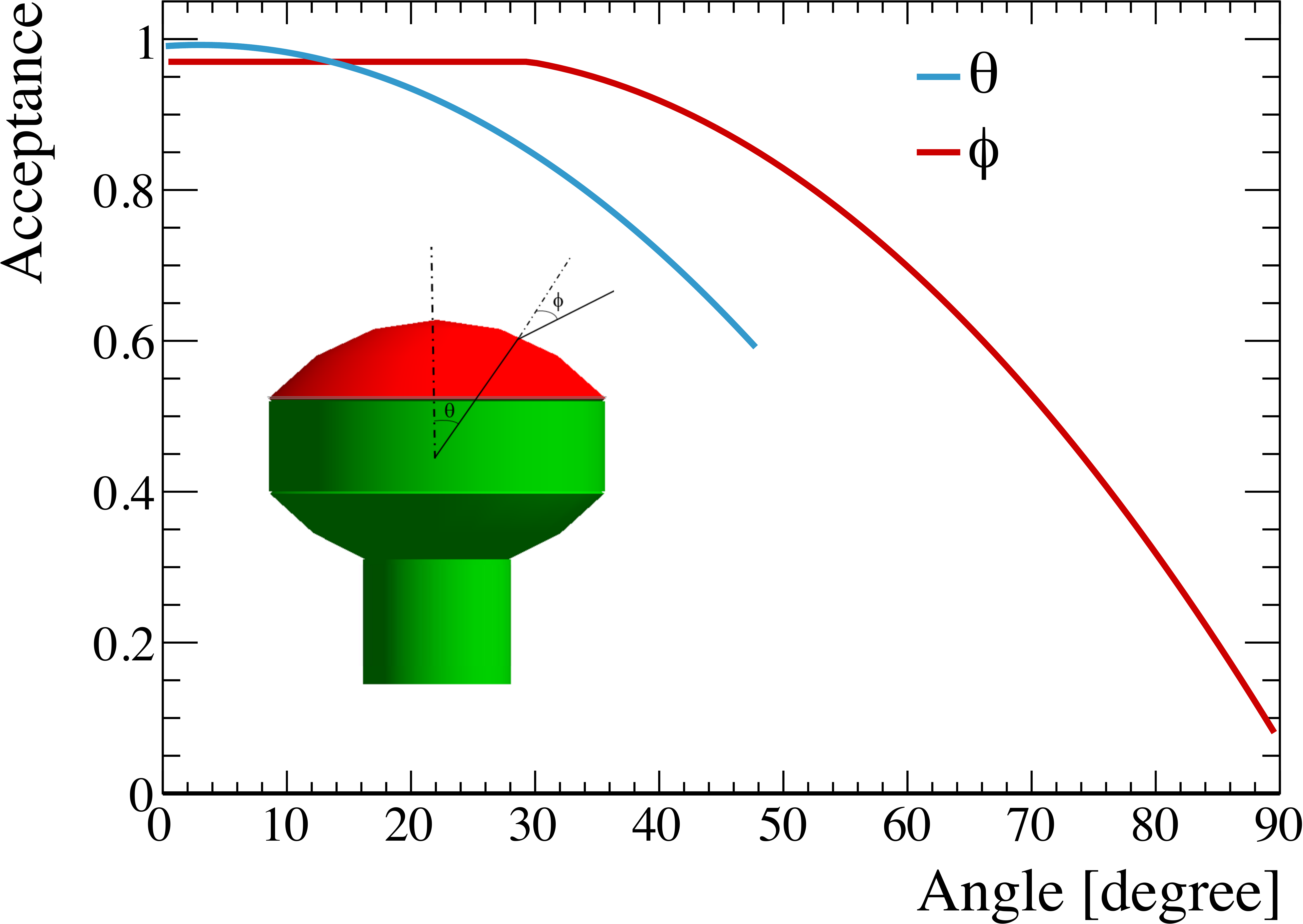}
    \caption{Left: Path of simulated VUV photons inside the \three simulation for a Rayleigh scattering length of \SI{20}{\centi\meter} (top) and \SI{163}{\centi\meter} (bottom). The implemented geometry can be seen: the CRP plane, the anode, the ground grid and the PMTs. For visibility, the field cage is not seen in these pictures. Right: Angular acceptance of photons on the photocathode.}
    \label{fig:simulation_param}
\end{figure}

A precise simulation of the scintillation light in the \three demonstrator is mandatory to study the parameters affecting the photon propagation and collection. 
On top of physics processes altering the photon propagation discussed in Sec.~\ref{sec1}, all main elements of the detector, such as the electric field cage, the cathode, the ground grid and the LEMs, are also influencing the amount of light to be detected.
Hence a detailed geometry of these detector elements has been implemented in \textsc{Geant 4}~\cite{Agostinelli:2002hh} as well as the PMTs, their photocathode designs, and the two TPB coatings. The implemented geometry can be seen in Fig.~\ref{fig:simulation_param}.\\

The Rayleigh scattering length for VUV photon in LAr is not well-known as its measurement is complicated due to high correlations with the LAr absorption length and the detector surface reflectivities.  Light simulations were generated with three different hypotheses of the Rayleigh scattering length: \SIlist{20;55;163}{\centi\meter}. 
While the two latter values are driven by the literature, the choice of the unrealistic shorter length was motivated by the sole purpose of quantifying its effect. 
Although a measurement of the Rayleigh scattering at \SI[separate-uncertainty=true,multi-part-units=single]{99.9+-0.8}{\centi\meter} has been recently performed~\cite{Babicz:2020den}, no simulations were produced at this value since the difference between $L_{Ray}=$\SI{55}{\centi\meter} and $L_{Ray}=$\SI{163}{\centi\meter} is tiny given the \three demonstrator dimensions, as it will be discussed in Sec.~\ref{sec6}. The Rayleigh scattering length for visible photons was set to \SI{350}{\centi\meter}~\cite{Seidel}. In Fig.~\ref{fig:simulation_param}-left, the path of photons simulated with the two extreme scattering values is shown.

The material reflectivity depends strongly on surface manufacturing. In our simulations, the default values are set to 0\% (50\%) reflectivity on stainless steel and copper for VUV (visible) photons respectively.

A 100\% TPB efficiency at \SI{128}{\nano\meter} is assumed and the visible photons are isotropically emitted by the TPB at wavelengths peaked at \SI{420}{\nano\meter}.
The PMT photocathode acceptance spans up to an angle of $\theta=\ang{48}$.
Effects of the photon impact position on the photocathode~\cite{1239351} and the photon incidence angle~\cite{hamamatsu} are taken into account following the parametrisation shown in Fig.~\ref{fig:simulation_param}-right.

\begin{figure}
    \centering
    \includegraphics[width=0.8\textwidth]{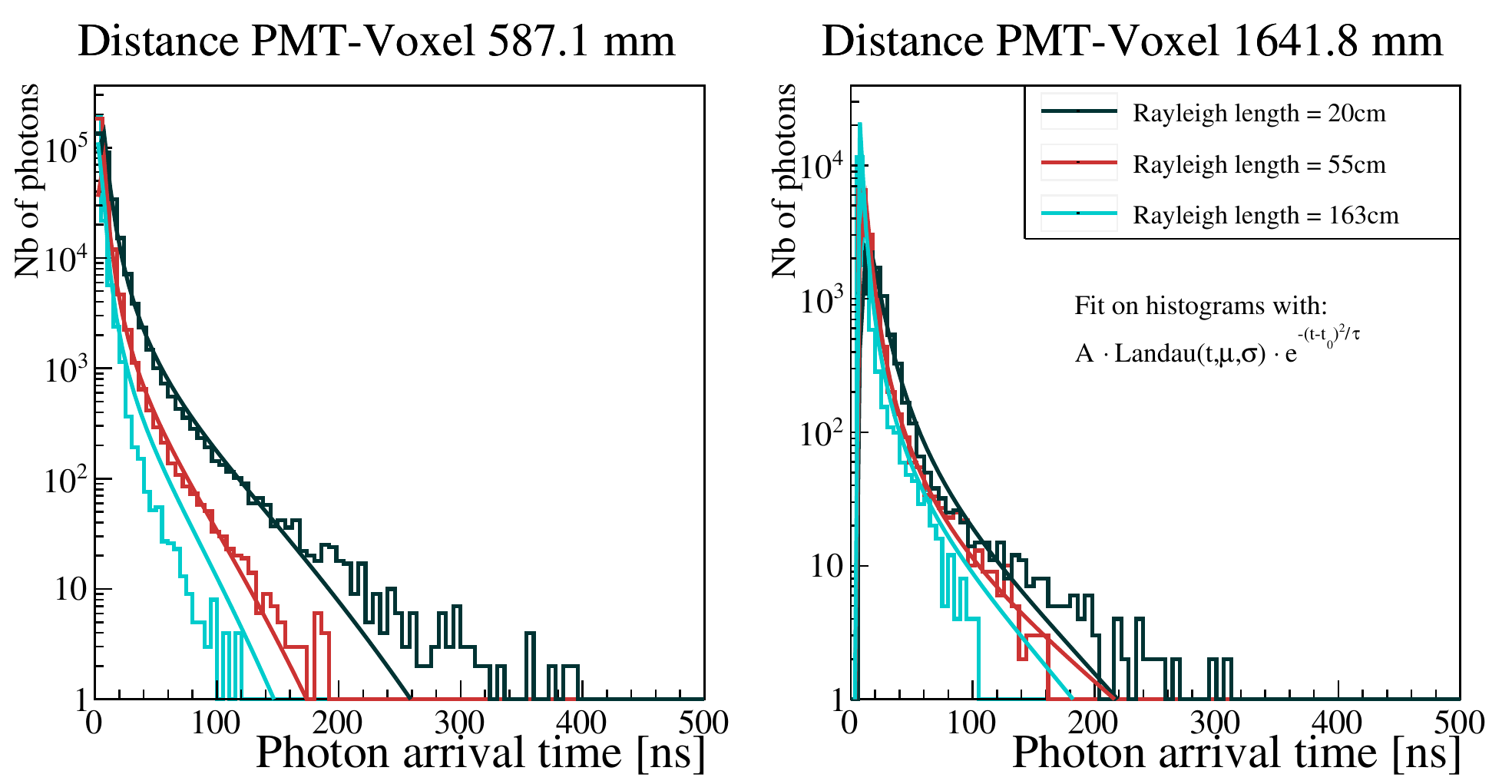}
    \caption{Arrival time distribution of photons for two different PMT-Voxel distances. Three values of the Rayleigh scattering length are assumed. The exponentially modified Landau fits are superimposed.}
    \label{fig:light_map_t_fit}
\end{figure}

\begin{figure}
    \centering
    \includegraphics[width=0.8\textwidth]{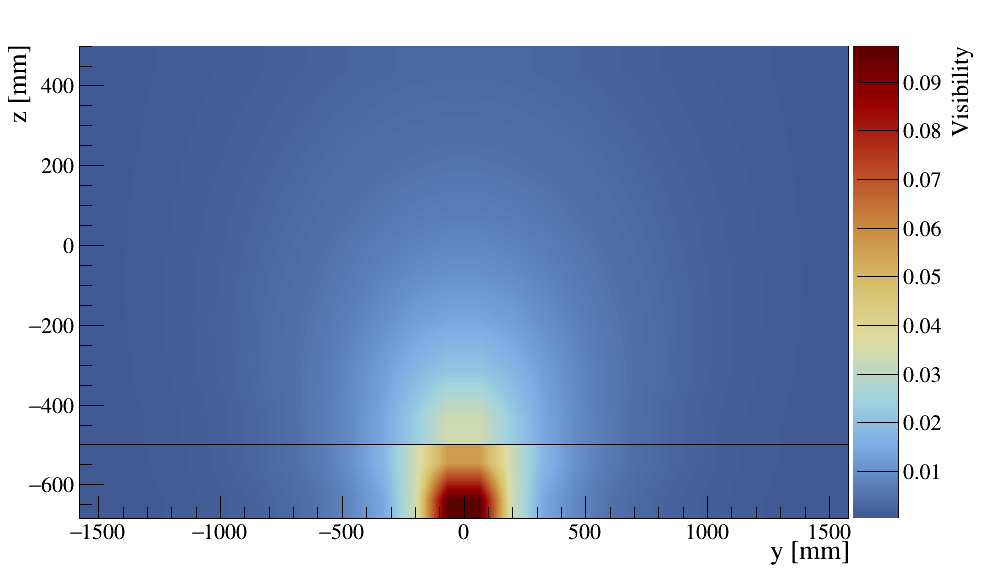}
    \caption{Interpolated visibility of PMT 3 along the $y-z$ plane at $x=0$~mm. The line at $z=\SI{-500}{\milli\meter}$ corresponds to the cathode. }
    \label{fig:light_map_w0}
\end{figure}

Given the very large number of scintillation photons produced per crossing particle -- about 10$^5$ photons/cm for a muon at the minimum ionizing potential in the absence of drift field -- and the collection efficiency of each PMT, a complete simulation of each photon trajectory would require a huge amount of CPU time.
An alternative solution has been developed to by-pass this problem: a generation of a so-called light map. 
The LAr volume is divided into 3D voxels\footnote{$\mathrm{N}_x\times\mathrm{N}_y\times\mathrm{N}_z = 24 \times 8 \times 8=1536$ voxels for the volume above the cathode, and $24 \times 8 \times 2=384$ below}. At the centre of every voxel, $10^8$ VUV photons are isotropically generated and propagated inside the detector. 
The number of detected photons and their arrival time is stored for each PMT-Voxel pair. An example of two of these distributions is shown in Fig.~\ref{fig:light_map_t_fit}. In order to keep the timing information for physics simulation, all distributions are adjusted with the same function and the extracted parameters are stored in the light maps. The distributions are influenced by the PMT-Voxel distance and by the value of the scattering length, making the choice of an appropriate function a complicated task. From similar studies made in a larger detector~\cite{anne_phd}, the Landau function has proven to give a satisfactory parametrisation of the photon arrival time distribution when PMT-Voxel distances are longer than a few meters. When the distance becomes shorter, the Landau tail gets more and more suppressed for late photons. To account for this feature in the relatively small size of this detector, the function used for the \three demonstrator is a Landau modified by an exponential term, overlaid in Fig.~\ref{fig:light_map_t_fit}. 
For each PMT-Voxel pair, five parameters are needed to retrieve the simulation output: $w_0$, the pair visibility defined as the ratio of collected over generated photons; $\mu$ and $\sigma$, the Landau parameters; $t_0$, the earliest photon arrival time and $\tau$, the exponential slope. These are the parameters stored in the light maps. 
The light maps are generated for a given scattering and reflectivity hypothesis to be used in physics simulations in place of the full propagation of each photon. 
During physics simulations, the number of photons produced at each step by either excitation or recombination is computed according to NEST predictions in liquid argon~\cite{Szydagis:2011tk}. 
The number of photons collected per PMT is extracted from the $w_0$ parameter stored in the light maps. The arrival time of each collected photons is determined by summing the track timing, the photon emission time according to its isospin state and the propagation delay. The latter is a random throw on the arrival time distribution retrieved from the light maps. A smooth variation of all stored parameters in the light maps in between nearby voxel centres is assumed. Due to discontinuities in the parameters at the level of the cathode, two light maps are needed for each simulation hypothesis.  In 
Fig.~\ref{fig:light_map_w0}, the 3D interpolated visibility for PMT 3 along the $y-z$ plane at $x=\SI{0}{\milli\meter}$ (centre of the detector) is shown for the light maps generated with $L_{ray}=\SI{55}{\centi\meter}$. 
The individual PMT quantum efficiency can be set at the physics simulation level. As this value is subject to large discussions among the LAr community, the default value used in the presented studies is set to 20\%, based on the manufacturer measurements~\cite{hamamatsu}. The light absorption length is set to infinity at the light map generation level. During the physics simulation, the effect induced by impurities on the amount of collected light can be approximated by introducing a $\exp({-L_{path}/L_{abs}})$ factor to the visibility, $L_{path}$ being the path length of the photon and $L_{abs}$ the absorption length.

\section{Scintillation light propagation measurements}
\label{sec6}
The mechanisms affecting the propagation of VUV scintillation light in liquid argon have been discussed in previous sections.
Given the size of the demonstrator, effects due to the liquid argon absorption length are considered to be negligible. 
The VUV photon reflection/absorption on the materials inside the TPC (stainless steel, copper) cannot be disentangled in such setup, and would require dedicated measurements. 
Depending on its value, the Rayleigh scattering length can have a quantitative impact on the amount of light collected.
Using a sample of muon-like tracks acquired with the shifted CRT trigger system, the evolution of the number of collected S1 signal as a function of the closest track-PMT distance, noted $\ell_{MA}$ in Fig.~\ref{fig:311diagram}, has been compared to Monte Carlo simulations generated with different scattering length hypotheses. 
Only runs taken with no drift field were considered for this analysis: the reconstructed S1 signal is not biased by a potential S2 signal. 
The CRT panels provide two 3D coordinates along the track path, from which it is possible to interpolate the closest point to each PMT. 
From all triggered events, a muon-like sample is selected by requesting that only one strip per CRT-panel is fired and that the measured time of flight across the cryostat is within $\pm\SI{40}{\nano\second}$. 
The reconstructed track path must cross the demonstrator active volume that is constrained by the field cage and stretches from the anode to the top of the PMTs. 
As the cathode introduces a discontinuity in the photon visibility, the closest track point to each PMT must be above that structure.
Finally, the selected events outside the linear PMT regime, as shown in Fig.~\ref{fig:satur} and~\ref{fig:comet} are discarded. About 30\% of the CRT triggered events satisfies these conditions. The S1 charge is integrated over [\SI{-40}{\nano\second};~+\SI{1000}{\nano\second}] around the trigger time.

In the simulation, \SI{4}{\giga\electronvolt/c} muons were generated from one CRT panel to the other following the ($\theta$, $\varphi$) space phase shown in Fig.~\ref{fig:crt-trigger} in the CRT-shifted case. The topological event selection and charge integration are the same as for the data. The light simulation was generated using the light maps produced with $L_{ray}=\;$\SIlist[list-units = single]{20;55;163}{\centi\meter} with full VUV photon absorption on stainless steel and copper.

\begin{figure}[h]
\begin{center}
\includegraphics[width=0.9\textwidth]{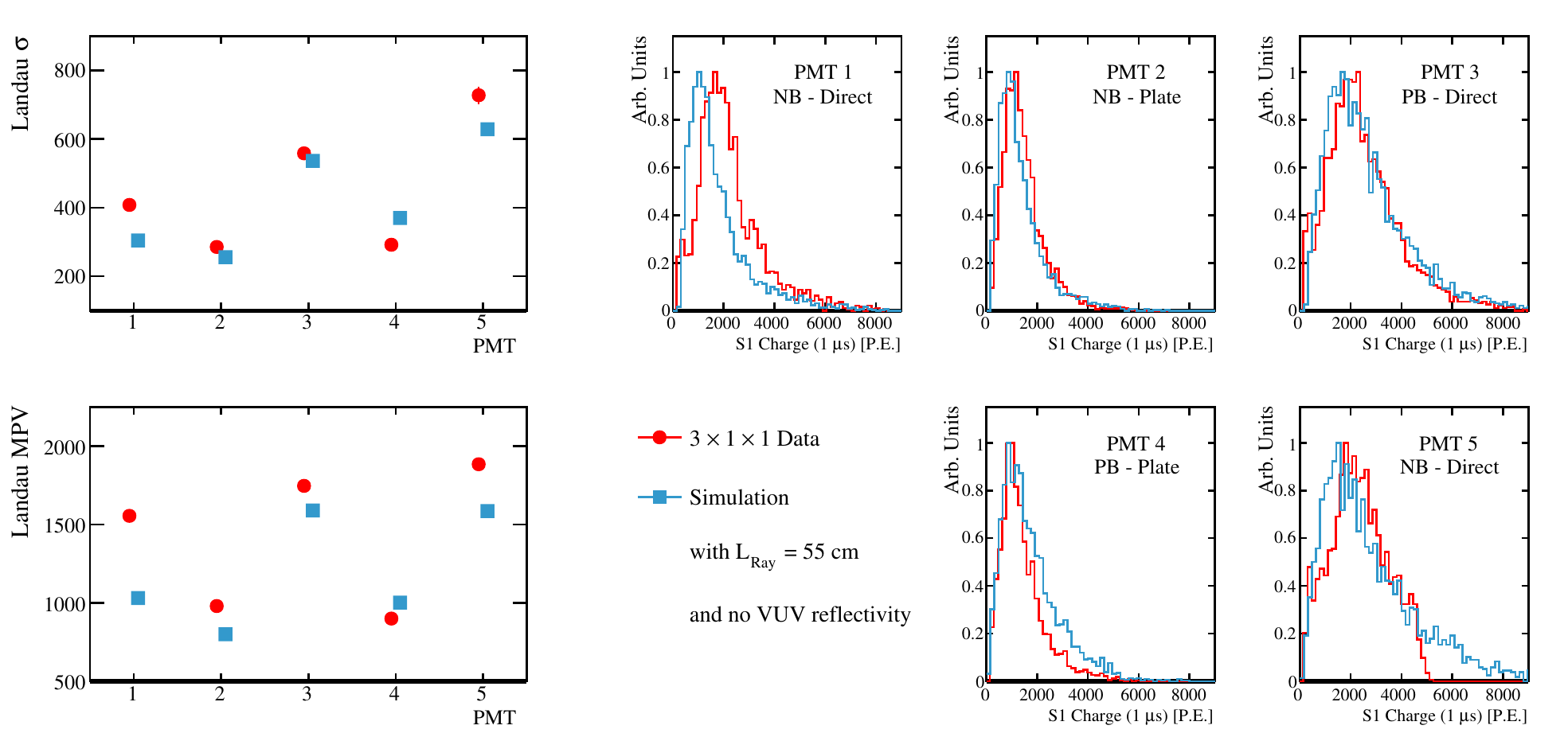}
 \caption{The S1 integrated charge over \SI{1}{\micro\second} of muon-like tracks in data (red) and in simulation (blue) are fitted with a Landau distribution, the extracted Most Probable Value (MPV) (left-bottom) and $\sigma$ (left-top) are shown. The distributions for all PMTs are presented on the right. In the data, only events in the PMT linear regime are kept, while this selection is not needed in the simulation.}
\label{fig:data_mc_comp}
\end{center}
\end{figure}

The validity of the simulation is presented in Fig.~\ref{fig:data_mc_comp} by comparing the integrated S1 charge over \SI{1}{\micro\second} with data for the selection of muon-like tracks explained previously. 
All distributions are fitted with a Landau function and the extracted parameters (most probable value and $\sigma$) are compared in the figure.
Due to the inclined topology of the tracks triggered by the CRT system, it is expected that the amount of light seen by PMT 1 is smaller than for PMT 5. 
This feature is well reproduced in the simulation. The PMTs having TPB coated on the PMMA plate (PMT 2 and 4) collect less light than the others, as seen in both data and MC distributions.
In terms of shape, the agreement between data and simulation appears to be fairly good. For the two PMTs located near the edge of the detector (PMT 1 and 5), the simulation predicts less light than what is measured in the data. This effect could be due to an underestimation of the stainless steel reflectivity for VUV photon -- set to 0\% by default in the simulation. 

\begin{figure}[h]
\begin{center}
\includegraphics[width=0.8\textwidth]{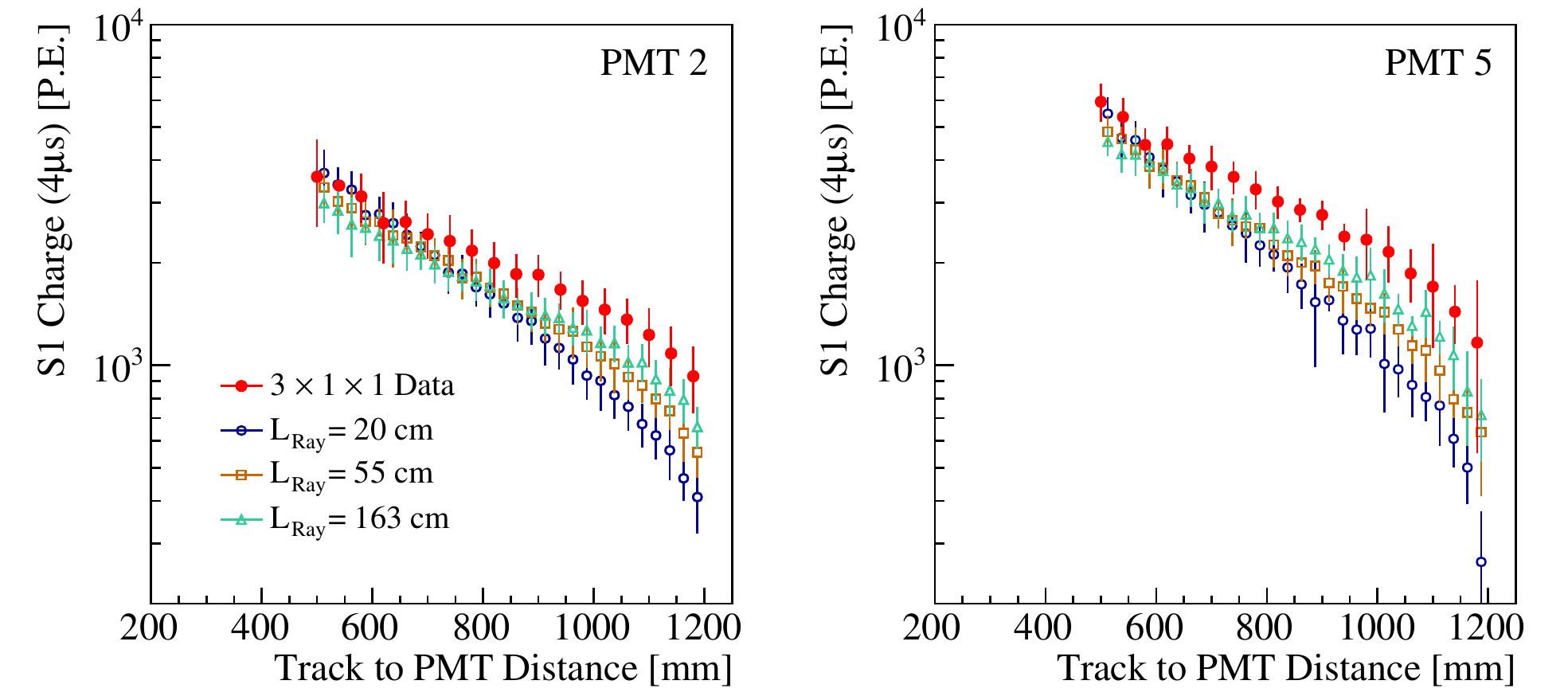}
\caption{Evolution of the S1 integrated charge over \SI{4}{\micro\second} as a function of the track to PMT distance for muon-like events collected at no drift field for PMTs 2 and 5. Each point is computed from a Gaussian fit of the distribution at a given distance. The error bars represent the $1\sigma$ spread. The data is presented in full red circle, the simulation assuming a Rayleigh scattering length of \SI{20}{\centi\meter} in open blue circle, \SI{55}{\centi\meter} in open orange square and \SI{163}{\centi\meter} in open green triangle. Total VUV photon absorption on the field cage, the cathode and the ground grid is assumed.}
\label{fig:rayleigh_data_mc}
\end{center}
\end{figure}
The amount of collected light as a function of the track to PMT distance is a handle to study the  Rayleigh scattering length. In Fig.~\ref{fig:rayleigh_data_mc}, the data is compared to the Monte Carlo predictions for two PMTs. 
First, one can note that the differences between the simulations and the data are contained within the error bars, so only a qualitative comparison can be performed. In the data, two regimes can be observed with a transition point at around \SI{600}{\milli\meter}. In the simulations, as the Rayleigh scattering length increases, this feature becomes more visible in agreement with the data. At long track to PMT distances, the simulations predict less light than seen in the data, which could be induced by an underestimated value of the material reflectivity.

\begin{figure}[ht]
\begin{center}
\includegraphics[width=0.8\textwidth]{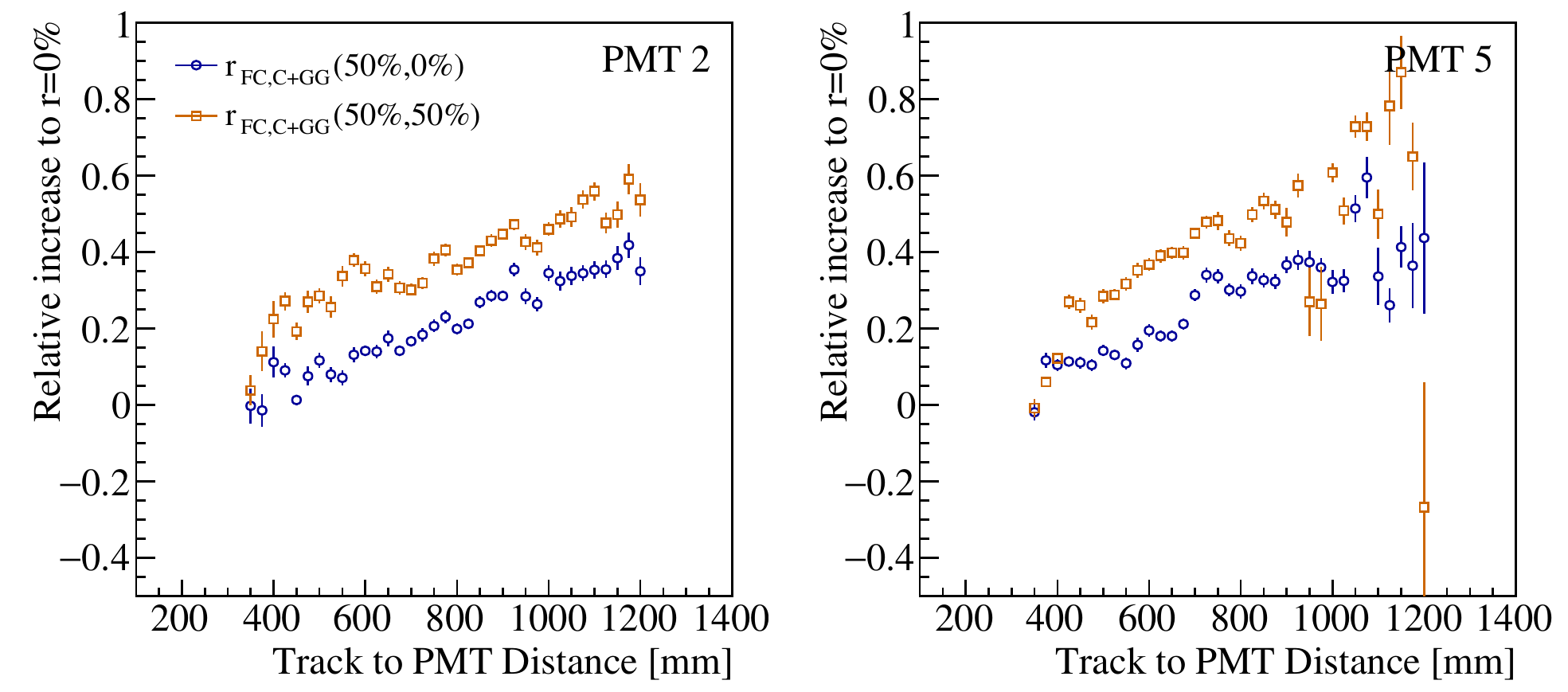}
\caption{Relative increase of the amount of light collected for PMTs 2 and 5 when different elements of the detector have a reflectivity set to 50\% for VUV photons. In blue circle, when only the field cage changes; in orange square, when all elements (field cage, cathode, ground grid) have a higher reflectivity. The Rayleigh scattering length is set to \SI{55}{\centi\meter}.}
\label{fig:reflection_mc}
\end{center}
\end{figure}

The effect of the VUV reflectivity on the amount of collected light is studied with simulations where the reflectivity on materials is increased to a larger value of 50\% at a fixed Rayleigh length of \SI{55}{\centi\meter}. Two simulations were performed: one where only the field cage reflectivity is increased, and one where all components have a 50\% reflectivity. The results are presented in Fig.~\ref{fig:reflection_mc} relative to the case with full VUV absorption on the detector components. The amount of collected light gradually increases with the track to PMT distance. At large distances, a $\sim50$\% increase of light is observed, where $\sim40$\% is attributed to the field cage reflection. From this study, it is clear that not only the amount of light changes with the reflectivity, but also the shape of the distribution.\\

These results are not to be interpreted as a measurement of the Rayleigh scattering length or the reflectivity. This is an indication of how these parameters affect the strength and the shape of the light curves. Additional factors as the effect of impurities are not considered in this study and could also play a role. 
Nevertheless, with this analysis technique, a larger detector should have a better sensitivity on both of these parameters as the efficiencies determined by the simulations diverge as the track-PMT distance increases.

\section{Scintillation light production measurements}
\label{sec7}

The measurement of the light yield as a function of the drift field allows to study the suppression of the electron-ion recombination process. 
The same selection of muon-like events triggered by the CRT system described in Sec.~\ref{sec6} is applied to data collected at the nominal drift field. In order to remove potential contributions of the S2 signal during the integration of the S1 charge, only events where the S2 would start after the end of the S1 integration range are considered.
As it will be discussed later in this section, our data suggests a variation of the fraction of fast and slow states with the drift field. 
Hence, for this study, the S1 charge integration range was increased up to \SI{3.5}{\micro\second} after the S1 peak to minimize the possible bias introduced by this effect. 
The drift field in the active volume is computed as the differential potential between the cathode and the first field shaper (FFS). Although the cathode voltage fluctuations are negligible in the runs used (below the per mille level), an uncertainty of 3.5\% is considered to account for the field cage ring diameter affecting the FFS-cathode distance.

\begin{figure}[ht]
\begin{center}
\includegraphics[width=0.42\textwidth]{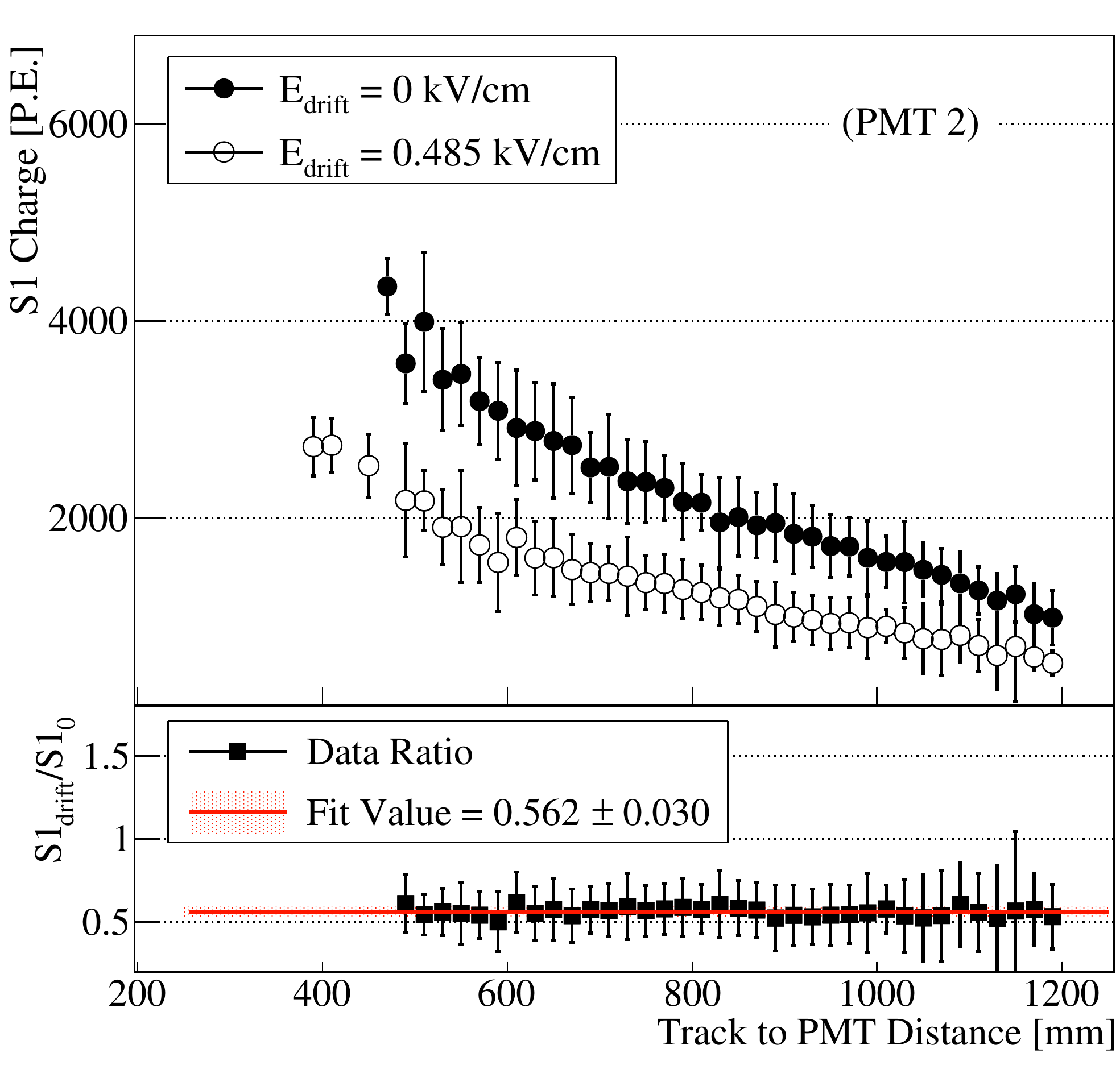}
\includegraphics[width=0.45\textwidth]{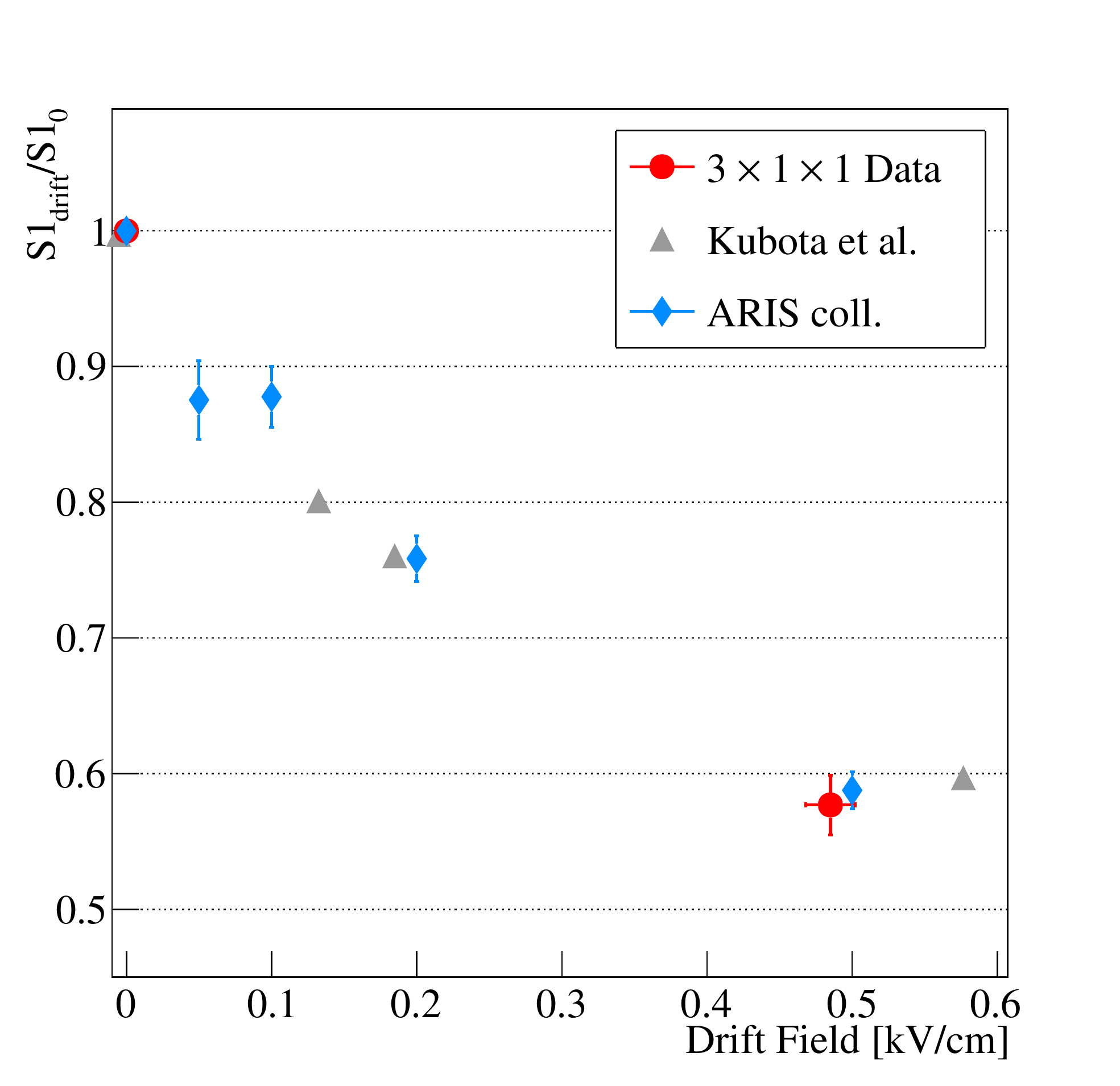}
\caption{Left: S1 charge as a function of the track to PMT distance with and without drift field and the ratio for PMT 2. Right: S1 charge as a function of the applied drift field normalized to the S1 charge with no drift field. The red point is the measurement of the \three demonstrator (all PMTs combined) compared to the values reported by ARIS~\cite{Aris} (light blue diamond) and Kubota \emph{et. al}~\cite{Kubota} (gray triangle).}
\label{fig:birks}
\end{center}
\end{figure}

Fig.~\ref{fig:birks}-left presents the reduction of the S1 charge as a function of track-PMT distance for data taken at two drift field strength. Combining the results of all PMTs, the reduction factor on the amount of light collected at the nominal drift field is $0.577\pm0.22$ with respect to data taken at no drift field. This highlights that without field, more than 40\% of the scintillation light is produced through the recombination process.
Our result is in agreement with previous studies done using electrons at the minimum ionizing potential~\cite{Kubota, Aris}, as shown in Fig.~\ref{fig:birks}-right. 

\begin{figure}[ht]
\begin{center}
\includegraphics[width=0.9\textwidth]{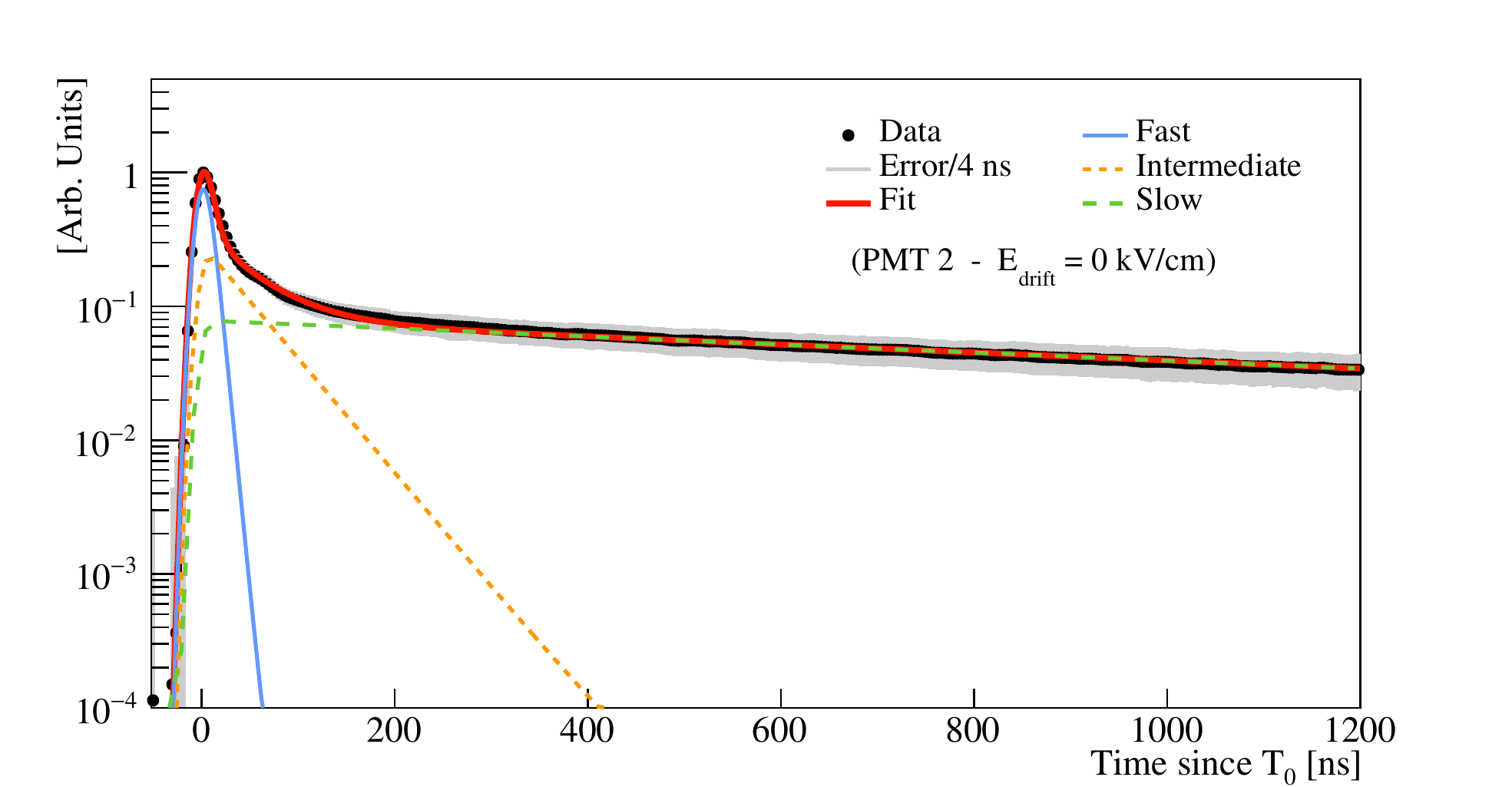}
\caption{Averaged normalized waveforms of muon-like events for one PMT without drift field (black points). The fitted function described in Eq.~\ref{eq:time} is overlaid in red. The individual contributions of the three components are also shown.} 
\label{fig:wf_fit}
\end{center}
\end{figure}

The scintillation light emission in LAr has a characteristic time dependence as mentioned in Sec.~\ref{sec1}. 
The amplitude-normalized averaged waveform of muon-like events tagged by the CRT system in the absence of drift field for one PMT is shown in Fig.~\ref{fig:wf_fit}. The waveforms can be described by a sum of exponentials to characterize the scintillation time, convoluted with a Gaussian to represent the detector response:
\begin{equation}
f(t) = \sum_{j=f,i,s} \frac{A_{j}}{2\tau_{j}}\exp{\left[\frac{1}{2}\left(\frac{\sigma}{\tau_{j}}\right)^{2}-\left(\frac{t-t_{m}}{\tau_{j}}\right)\right]}\left[1-{\rm erf}\left(\frac{1}{\sqrt{2}}\left(\frac{\sigma}{\tau_{j}}-\frac{(t-t_{m})}{\sigma}\right)\right)\right]. 
\label{eq:time}
\end{equation}
Although the scintillation time profile should have only two components, from the decay to ground state of singlet (fast) and triplet (slow) argon excimers, a third component -- later referred to as the intermediate one -- has to be added in order to allow the fit to converge. Most of the LAr based experiments have also reported this third time structure, with yet no clear consensus on its origin. Typical values reported lay in the range of $\tau_i\sim\;$\SIrange{50}{130}{\nano\second}. In~\cite{Segreto, Whittington}, the hypothesis of a delayed emission time by the wavelength shifting material is explored.  In Eq.~\ref{eq:time}, $j$ refers to the three exponential components ($f$ fast, $i$ intermediate, and $s$ slow), $A_j$ are normalization constants, $t_{m}$ and $\sigma$ are the Gaussian mean and width respectively and $\tau_j$ are the parameter characterizing the exponential fall-offs. In Fig.~\ref{fig:wf_fit}, the fitted function on the averaged waveforms is superimposed. Given the digitization sampling of \SI{4}{\nano\second}, the fit has a limited sensitivity to $\tau_f$: this parameter is fixed to \SI{6}{\nano\second}~\cite{Hitachi}. 
The function of Eq.~\ref{eq:time} cannot describe correctly PB PMTs waveforms due to the presence of the signal reflections. For the rest of this section, only NB PMTs are analyzed.

As discussed in Sec.~\ref{sec1}, quenching processes due to the presence of impurities (like O$_{2}$ and N$_{2}$) can modify the signal time profile by decreasing the slow component lifetime and relative amount proportionally to their concentrations.
Hence, a deterioration of the LAr purity can be detected by monitoring the lifetime of the slow component. From literature, this method is sensitive to concentrations down to \SI{0.1}{ppm}, \SI{3}{ppm} and \SI{10}{ppb} of O$_2$, N$_2$ and CH$_4$ respectively. 
At the end of the cooling down of the cryostat, the impurities were measured to be at \SI{0.15}{ppm}, \SI{50}{ppb} and \SI{7}{ppm} for O$_{2}$, N$_{2}$, and moisture respectively using residual gas trace analysers~\cite{311}. 

\begin{figure}[ht]
\begin{center}
\includegraphics[width=0.95\textwidth]{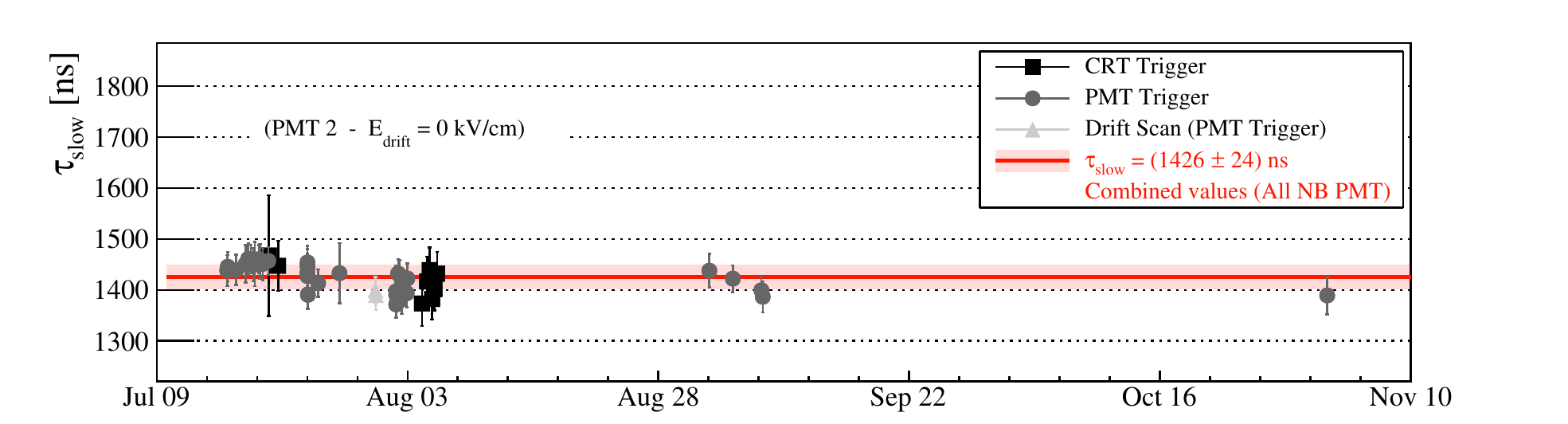}
\caption{Evolution of $\tau_{s}$ for one PMT as a function of time once the cryostat is filled with LAr. Only runs taken at no drift field are considered.  The combined value for all NB PMTs is superimposed in red.}
\label{fig:moni_tauslow}
\end{center}
\end{figure}

For this analysis, only runs taken without drift field are considered. This is justified by the observation of the evolution of the $\tau_{s}$ lifetime with the drift field, that will be discussed later. In total, 48 runs have been considered. 
For CRT trigger runs, the muon-like selection previously described is applied. For PMT self trigger runs, all events lying in the linear regimes are considered. The obtained averaged waveform for each run and each NB PMT is fitted with the Eq~\ref{eq:time}.
Fig.~\ref{fig:moni_tauslow} shows the extracted $\tau_{s}$ for one NB PMT, and the results are similar in the other two. 
The obtained values are stable from July to November, and do not depend on the event selection driven by the trigger condition. Combining all NB PMT results, a constant value of $\tau_{s}=\SI[separate-uncertainty=true,multi-part-units=single]{1426+-24}{\nano\second}$ is found, which is compatible with the low contamination levels measured at the beginning of the detector operation and with the measurement of the electron lifetime performed in the demonstrator~\cite{311}. 

The relative population of the singlet and triplet excited states has been shown to be dependent on the deposited energy~\cite{Hitachi}. In a paper from Kubota {\it et al.}~\cite{Kubota} using $\mathcal{O}$(MeV) electrons, the population of these two states produced either through recombination or excitation has been measured to be different: 
$$\frac{A_f}{A_s}[\mathrm{Rec}]=0.5\pm0.2\;\;\;\;\frac{A_f}{A_s}[\mathrm{Exc}]=0.36\pm0.06.$$
According to this observation, the $A_f/A_s$ ratio is then expected to decrease with the drift field, as the light emitted by electron-ion recombination gets suppressed.

\begin{figure}[ht]
\begin{center}
\includegraphics[width=0.49\textwidth]{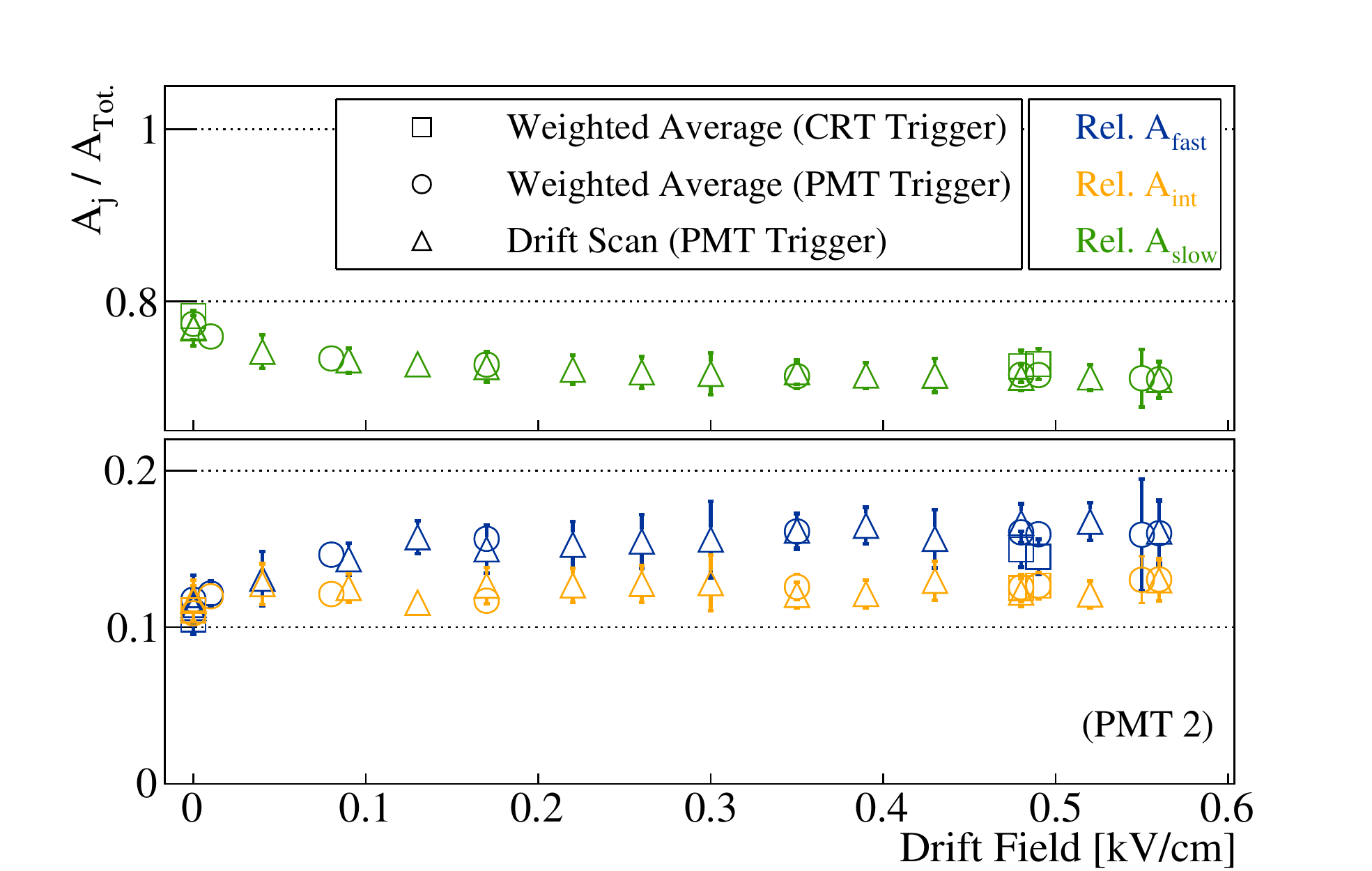}
\includegraphics[width=0.49\textwidth]{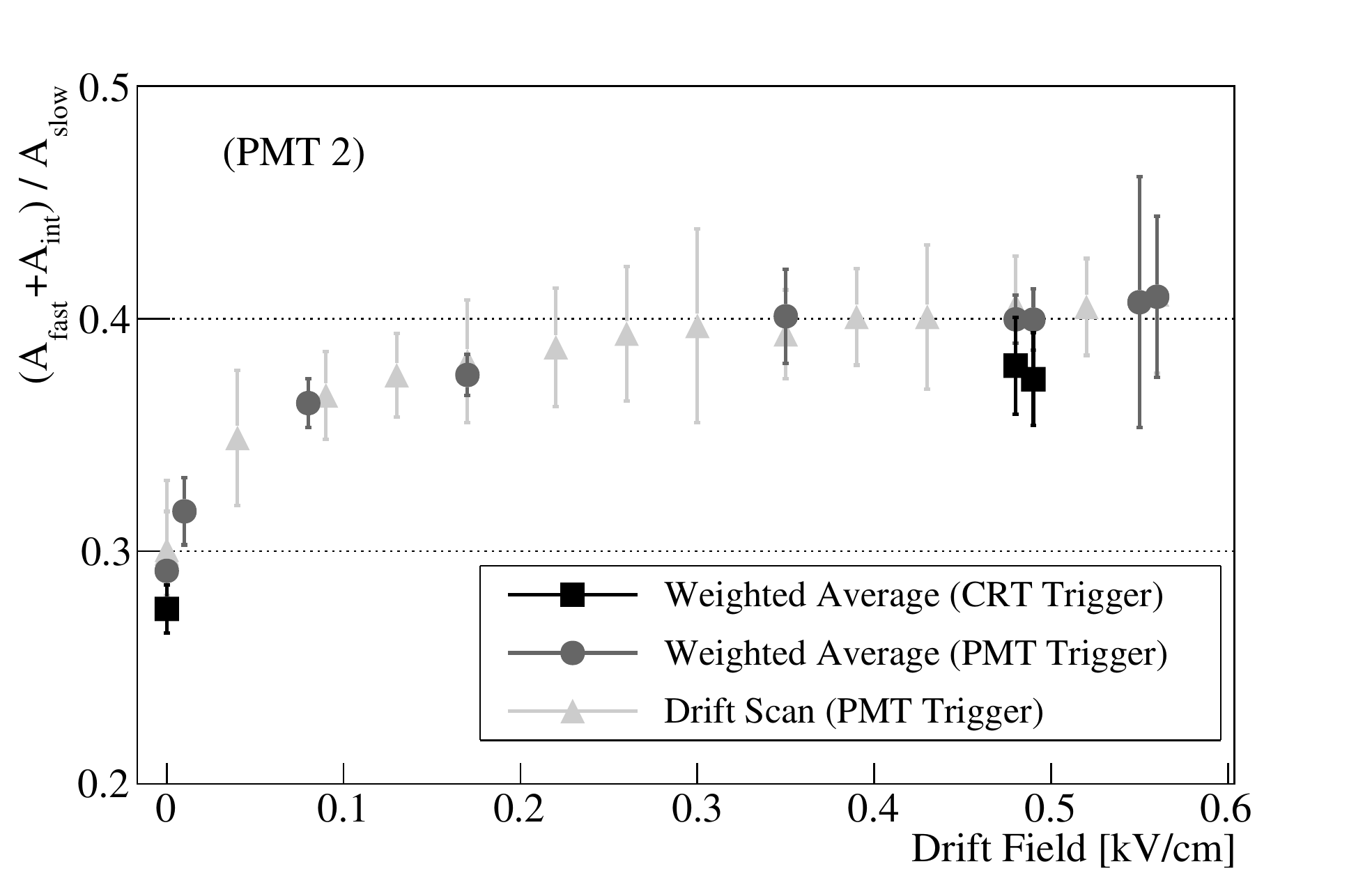}
\caption{Left: Relative normalization constants $A_{j}/(A_{f} + A_{i} + A_{s})$ ($j=f,i,s$) as a function of the applied drift field for PMT 2. Right: $(A_{f} + A_{i})/A_{s}$ versus the drift field for the same PMT. Results are similar for the other two NB PMTs.}
\label{fig:aratio}
\end{center}
\end{figure}

This analysis is carried out using the light data collected in the \three, using the same selections described previously for CRT trigger and PMT self trigger runs. Although having low statistics, the drift scan runs have been used here considering only events in the PMT linear regime.
The averaged waveforms are fitted for each run and each NB PMT with the Eq.~\ref{eq:time}. The extracted normalization constants are reported in Fig.~\ref{fig:aratio}-left for one PMT. For clarity, results from runs with the same trigger condition and drift field, initially fitted separately, have been combined together in the figure. 
While the relative contributions of fast and slow components are affected by the presence of the drift field, the intermediate component contribution, $A_{i}$, remains constant. Fig.~\ref{fig:aratio}-right shows the evolution of the observable $(A_{f} + A_{i})/A_{s}$ with the drift field. The intermediate component has been treated together with the fast one here, but the global behavior is independent from this choice. In~\cite{Acciarri}, a variation of singlet and triplet states with N$_2$ concentration has been measured. This explanation is discarded here as the $\tau_{s}$ has been shown to be stable over time, and by the fact that the trend is seen with the drift scan runs taken over a span of a couple of hours. The extracted values from muon-like samples (CRT trigger) and cosmic rays (PMT self triggers) agrees within the error bars. The obtained values cannot be directly compared to previous results, as the treatment of the intermediate part is different -- or not clearly explained -- and by the fact that the fitting function and procedure is not always the same as ours. 
In particular, an opposite trend is seen from Kubota's results, although the analysis presents several differences.
In~\cite{Kubota}, scintillation time profile taken at a drift field of \SI{6}{\kilo\volt/\centi\meter} is assumed to be driven by the excitation process only, while the curve difference between data taken at \SI{0}{\kilo\volt/\centi\meter} and \SI{6}{\kilo\volt/\centi\meter} extracts the recombination process. Although the intermediate component is seen, it is not clear in \cite{Kubota} how it is treated in the reported values of $A_{f}/A_{s}$. Moreover, the quoted LAr purity of their apparatus falls in the range where the time profile gets affected.

Using the \three data, an increase in ($A_f+A_i)/A_s$ of 34$\%$ is observed at \SI{0.5}{\kilo\volt/\centi\meter} with respect to the no drift field case. This result suggests that the relative population of singlet and triplet states created through the recombination is lower than through direct excitation: $\frac{A_f}{A_s}[\mathrm{Rec}] < \frac{A_f}{A_s}[\mathrm{Exc}]$. To our knowledge, it is the first time that the evolution of the relative population of singlet and triplet states is measured at several drift field settings in LAr. 
In many LArTPC experiments, the particle identification is obtained by pulse shape discrimination (PSD) of the S1 peak -- in particular to discriminate electronic from nuclear recoils in Dark Matter experiments. As the PSD is driven by the $A_f/A_s$ ratio, more dedicated studies of this ratio evolution with the drift field, the purity and particle type should be considered by the LAr community.

\begin{figure}[ht]
\begin{center}
\includegraphics[width=0.49\textwidth]{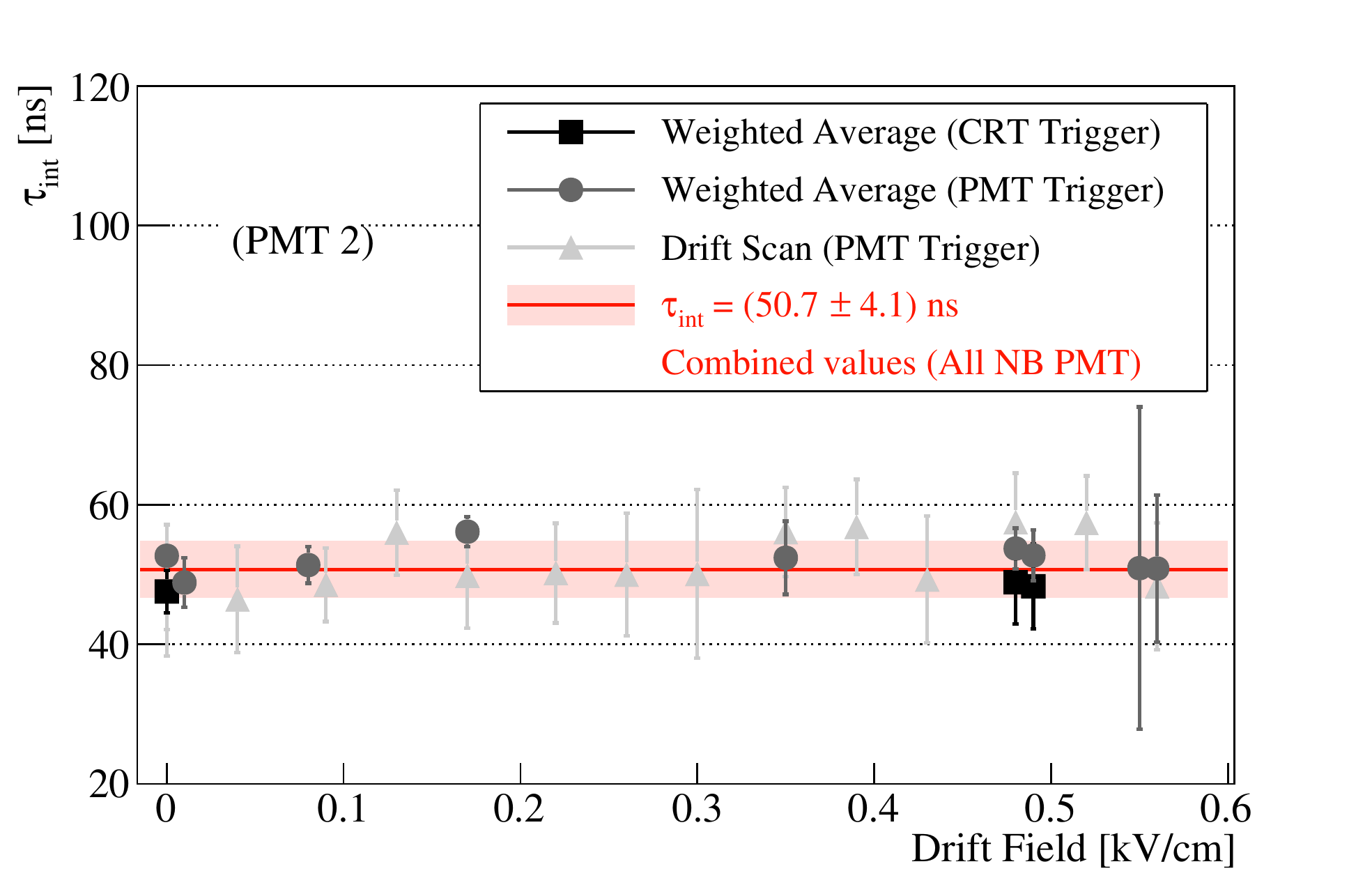}
\includegraphics[width=0.49\textwidth]{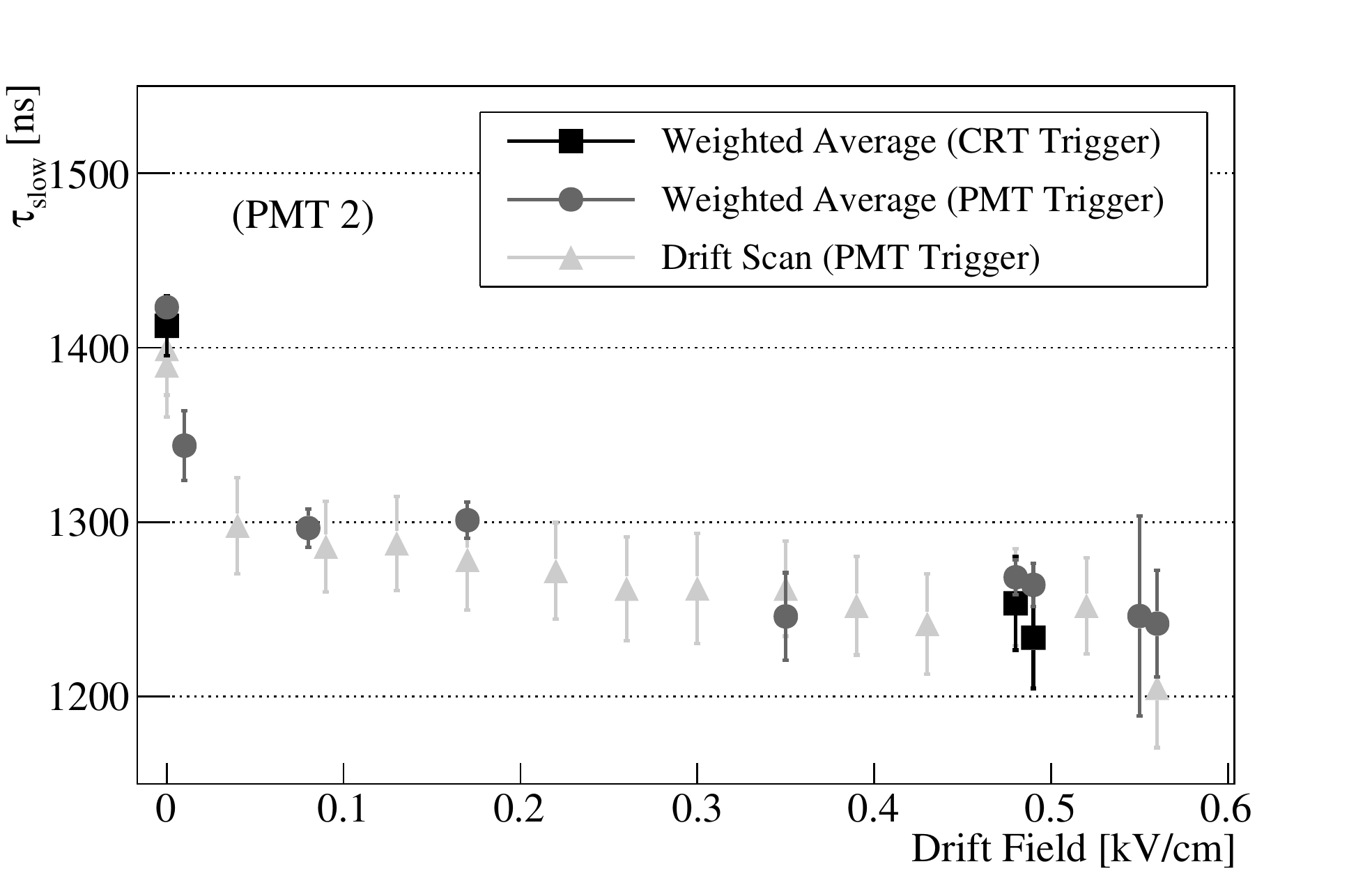}
\caption{Evolution of $\tau_{i}$ (left) and $\tau_{s}$ (right) with the drift field for PMT 2. Results are similar for the other two NB PMTs. 
}
\label{fig:slowvsE}
\end{center}
\end{figure}

Using the same runs, selections and fitting procedure, the evolution of the intermediate and slow lifetimes with the drift field has been performed. In Fig.~\ref{fig:slowvsE}-left, the intermediate decay time is shown. A value of $\tau_{i}=\SI[separate-uncertainty=true,multi-part-units=single]{50.7+-4.1}{\nano\second}$ is found independently of the field. 
On the other hand, $\tau_{s}$, shown in Fig.~\ref{fig:slowvsE}-right, decreases with the drift field. This behavior cannot be attributed to a variation of the impurities. At a drift field of \SI{0.5}{\kilo\volt/\centi\meter}, the combined value is 10\% lower than without field: $\tau_{s}= \SI[separate-uncertainty=true,multi-part-units=single]{1278\pm13}{\nano\second}$. To our knowledge, a campaign of measurements on the evolution of $\tau_{s}$ with the drift field has never been reported before in LAr. For completeness, a lower value of  $\tau_{s}$ for a very strong drift field of \SI{6}{\kilo\volt/\centi\meter} is mentioned in \cite{Kubota}; however, this analysis lacks of details concerning the LAr purity stability over different measurements. 
A similar trend has been observed in LXe \cite{Kubota, Hogenbirk}, interpreted as being the consequence of a recombination time being not negligible compared to the singlet and triplet decay times. In LAr, the recombination time has been measured: $T_r=\SI[separate-uncertainty=true,multi-part-units=single]{0.8\pm0.2}{\nano\second}$~\cite{Kubota}, which is too fast to have a significant impact on the scintillation time profile.

\section{Study of the secondary scintillation signal}
\label{sec8}

\begin{figure}[ht]
\centering
\includegraphics[width=\textwidth]{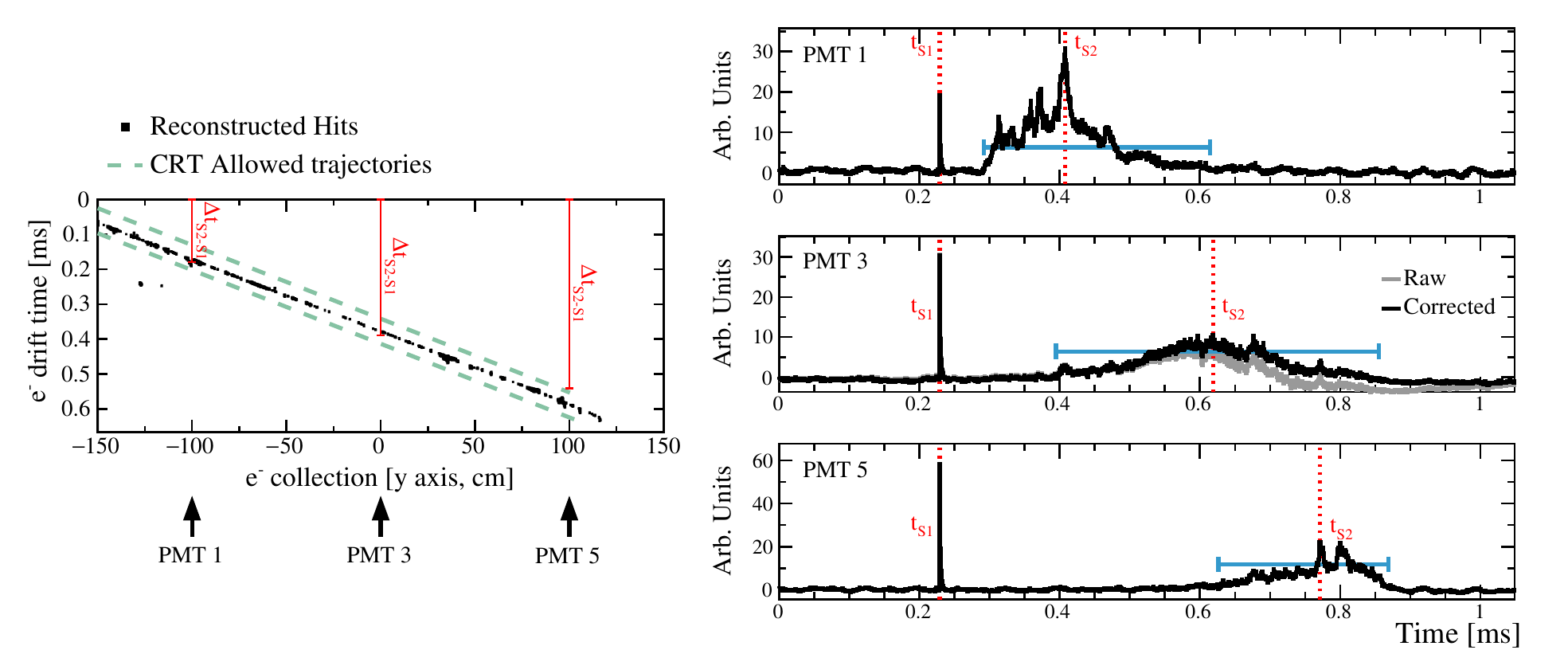}

\caption{Example of event taken with the \three demonstrator with CRT trigger. The reconstructed charge hits along the $y-z$ view are shown on the left, as well as the CRT reconstructed track paths. The light collected in PMT 1, 3 and 5 are shown on the right. The correction for the capacitance overshooting in PMT 3 (positive base) is shown. Reconstructed S1 and S2 times are shown in red. The reconstructed duration of S2 signal is shown in blue.}
\label{fig:event}
\end{figure}

The extraction of the drifted electrons in the gas phase produces a secondary scintillation (S2) signal proportional to the drifted charge. 
The time extension of the S2 signal is related to the track inclination with respect to the CRP. The more inclined the track is, the larger the S2 time extension is. The time at which S2 is maximal corresponds to the arrival time of the electrons produced above the PMT inside the LEMs. The difference between this time and the S1 maximum corresponds to the drift time of the electrons from the point of interaction to the anode. A dedicated algorithm, validated on MC simulations, has been developed to extract the S2 starting, maximal and ending times as well as the S2 highest amplitude and integrated charge. In Fig.~\ref{fig:event}, S2 waveforms and reconstructed times are shown. 

If the drift distance is known, the electron drift velocity can be measured using the time difference between S1 and S2 highest signals. In the demonstrator, two analyses were performed under this purpose. 

\begin{figure}[ht]
\begin{center}
\includegraphics[width=0.45\textwidth]{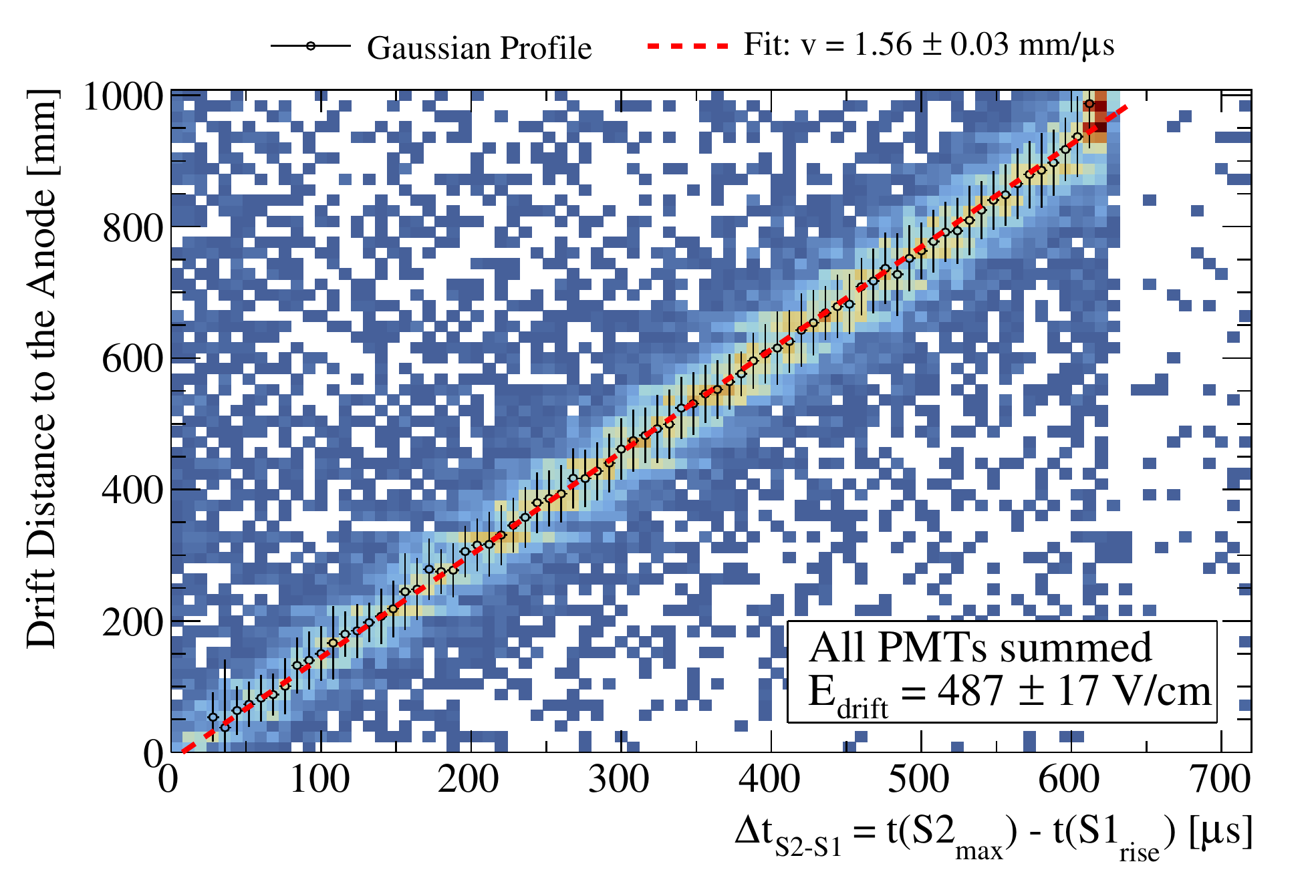}
\includegraphics[width=0.45\textwidth]{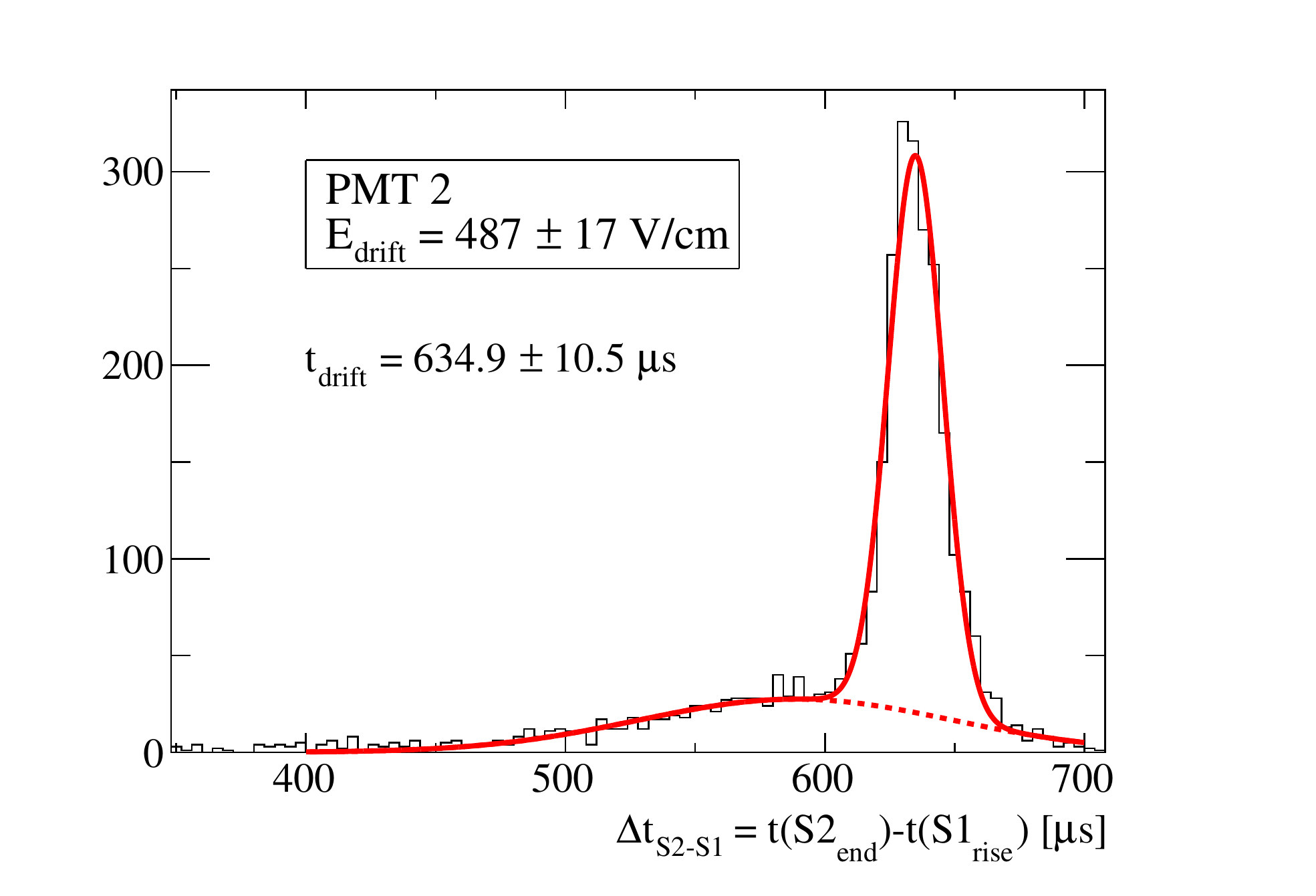}
\caption{Left: From the CRT-based analysis, the relation between the e$^-$ drift distance to the anode and $\Delta t_{S2-S1}$ time. The profile distribution, computed by a Gaussian fit of each $\Delta t$ slice, is overlaid as well as the linear fit in red. Right: From the PMT-based analysis, the S2 ending time distribution for muon-like tracks escaping the detector through the cathode. In red, the two Gaussian fits (for the continuum and the peak) is shown.}
\label{fig:ana_vel}
\end{center}
\end{figure}

For runs taken in CRT-trigger mode, after the selection of muon-like tracks described in Sec.~\ref{sec6}, one can compute the drift distance above each PMT. In Fig.~\ref{fig:ana_vel}-left, the 2D histogram of the drift distance versus the drift time, $\Delta t_{S2-S1}$, is shown as well as the linear drift velocity fit. As the CRT strips have a width of \SI{11}{\centi\meter}, the position of the track inside the fiducial volume comes with an uncertainty affecting the drift velocity results. This systematic error, of about 2\%, is added in quadrature with the error from the fit. 

For runs taken in PMT-trigger mode, charge and light events were matched together offline based on the event timing (both with a ms precision) comparison. A set of algorithms has been used for the track reconstruction: noise filtering, hit finder, 2D clustering and finally 3D merger ~\cite{andrea-thesis}. From all reconstructed tracks, a selection of crossing muon-like events was performed considering mainly the track topology -- an entry and exit points inside the fiducial volume must be found -- and the transverse spread of charge -- to remove electromagnetic showers mis-reconstructed as straight lines. Events containing tracks arriving before or after the triggered one are discarded in order not to bias the S2 reconstructed variables. As the charge reconstruction provides the position and the arrival time of the electrons at the anode, we cannot perform a similar analysis as for the CRT-trigger runs. Instead, from the reconstructed topology of tracks, selecting triggering muons crossing the entire drift volume (from the anode to the cathode) is possible. Therefore, as the exiting point of the track is a known position inside the detector, the S2 ending time provides the drift time of the electrons over the maximal \SI{1}{\meter} drift distance. In Fig.~\ref{fig:ana_vel}-right, the S2 ending time distribution for one PMT is presented and fitted with two Gaussians, to represent the peak and the continuum. The results from all PMTs, consistent within each other, are combined together.

\begin{figure}[ht]
\begin{center}
\includegraphics[width=0.9\textwidth]{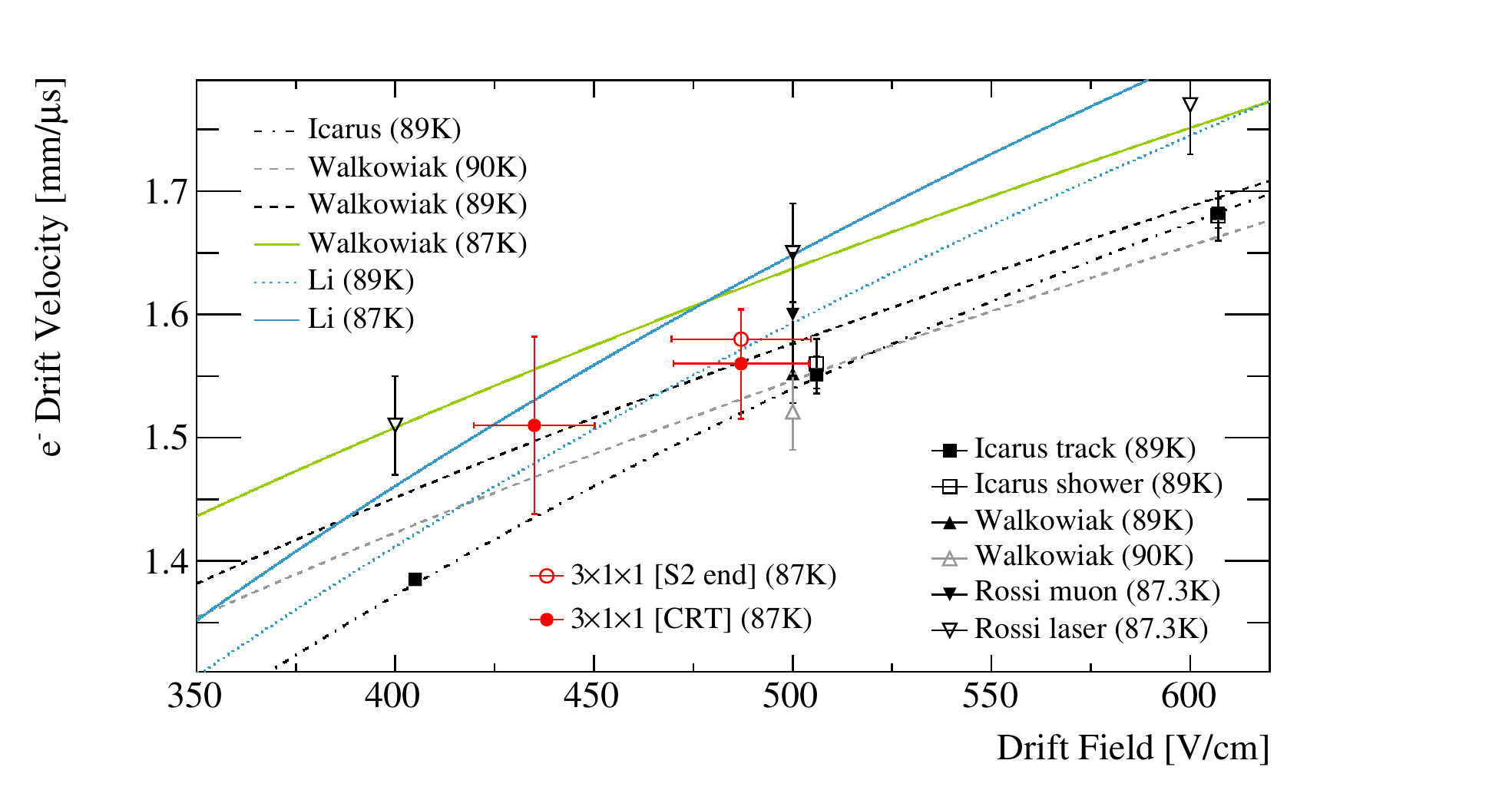}
\caption{Measurement of the electron drift velocity in the \three demonstrator compared to data and  parametrisations of ICARUS~\cite{Amoruso:2004ti}, Walkowiak~\cite{Walkowiak:2000wf}, Li {\it et al.}~\cite{Li:2015rqa} and Rossi {\it et al.}~\cite{Rossi:2009im} at different LAr temperatures.}
\label{fig:drift_vel}
\end{center}
\end{figure}

In Fig.~\ref{fig:drift_vel}, the results from the \three demonstrator are summarized and compared to different drift velocity parametrisations and measurements from literature. The measurement was performed only at two drift fields, where runs were taken with a long data acquisition window and sufficient statistics. For this analysis, we assumed no space-charge effect that would distort the drift field and bias the drift velocity measurement.
Finally, we did not take into account the fact that electrons will experience different electric fields during their journey to the anode, in particular in the last 13~mm before collection, as seen in Fig.~\ref{fig:DP_schema_311_pic}. At the nominal drift field, the two analyses give consistent results: $\varv[\mathrm{S2 end}] = \SI[separate-uncertainty=true,multi-part-units=single]{1.58\pm0.02}{\milli\meter/\micro\second}$, $\varv[\mathrm{CRT}] = \SI[separate-uncertainty=true,multi-part-units=single]{1.56\pm0.03}{\milli\meter/\micro\second}$ and are in agreement with the two parametrisations at T$_{LAr} =\SI{87}{\kelvin}$.

\begin{figure}[ht]
\begin{center}
\includegraphics[width=0.9\textwidth]{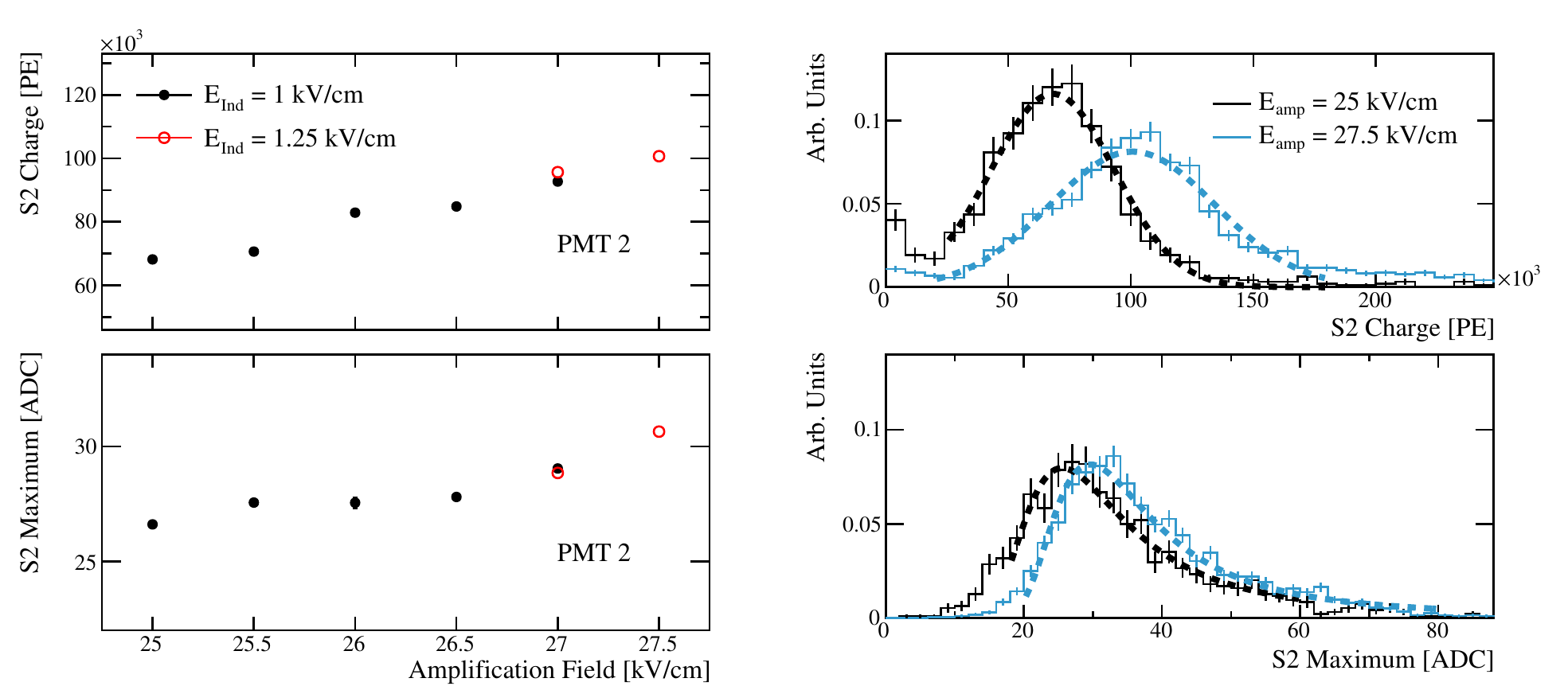}
\caption{Left: Reconstructed S2 charge (top) and highest amplitude (bottom) as a function of the amplification field for PMT 2 at two induction fields. The statistical errors are smaller than the markers. Right: Corresponding distributions (histogram) and fit (dashed line) for the lowest and highest amplification fields.}
\label{fig:S2_lem_field}
\end{center}
\end{figure}

Using the same muon-like selection in PMT- and CRT-trigger runs, we studied the evolution of the S2 maximal amplitude and total charge as a function of the amplification field. With the criteria of sufficiently long runs taken with a 1~ms acquisition window, similar drift and extraction fields and a uniform differential potential applied across all LEMs, a scan of the amplification fields between \SI{25}{\kilo\volt/\centi\meter} and \SI{27.5}{\kilo\volt/\centi\meter} in steps of \SI{0.5}{\kilo\volt/\centi\meter} is performed. In the runs selected, the drift field is \SI{0.49}{\kilo\volt/\centi\meter} and the extraction field in liquid is \SI{2.1}{\kilo\volt/\centi\meter} (\SI{3.2}{\kilo\volt/\centi\meter} in gas). In principle, the induction field should play a negligible role in the amount of S2 signal collected by the PMTs, due to the opacity of the LEMs. In the runs selected for this analysis, the induction field is at \SI{1}{\kilo\volt/\centi\meter} and \SI{1.25}{\kilo\volt/\centi\meter}. From the muon-like samples, the distributions of S2 maximal amplitude and charge are fitted with a Landau and a Gaussian function respectively. The extracted values are reported in Fig.~\ref{fig:S2_lem_field} for one PMT.
The results are the same for the other PMTs, confirming that our waveform overshooting correction procedure for positively biased PMTs is correct. The small increase of the induction field has no visible effect on the amount of S2 signal. The S2 maximal amplitude and charge increase with the amplification field. 
Between data collected at an amplification field of \SI{25}{\kilo\volt/\centi\meter} and \SI{27.5}{\kilo\volt/\centi\meter}, an increase factor of $\sim$1.5 for the S2 charge is measured whereas the S2 maximal amplitude only increases by a factor of $\sim$1.2.

These results might be explained at least qualitatively by the results of the simulation study presented in \cite{Lux_2019} using the same readout geometry although using a higher induction field of \SI{5}{\kilo\volt/\centi\meter}. 
Following that study, the charge gain is expected to increase by a factor of 1.5 to 2 when increasing the amplification field from \SI{25}{\kilo\volt/\centi\meter} to \SI{27.5}{\kilo\volt/\centi\meter}. 
This is in agreement with the measured charge gain increase shown in Fig.~\ref{fig:S2_lem_field}. The study from \cite{Lux_2019} also simulates the light production in the 
amplification region of a dual phase TPC. The S2 photons are mainly produced in the LEM holes and only the photons emitted within a small solid angle will leave the holes towards the PMTs. The S2 photons production point is, on average, closer to the anode as the LEM voltage increases. 
Therefore, the solid angle is not a constant but becomes smaller for higher amplification fields. 
This will affect the observed S2 light in two ways: on one hand, while the total S2 gain increases exponentially, the fraction of light reaching the liquid surface increases more slowly due to the solid angle reduction.
This effect might be enhanced due to the detector geometry. On the other hand, the shifting of the average production point of the S2 light also affects the angular distribution of the photons
going backwards to the PMTs, becoming narrower for larger LEM voltages. Since the five PMTs are installed underneath the junction 
of four LEMs and not below their center, this could reduce further the fraction of photons reaching the PMTs.

\section{Conclusions}

The performance of a photon detection system for a dual phase LArTPC with charge readout has been studied. In the 4~tonne demonstrator, five Hamamatsu R5912-02MOD PMTs have been installed with two configurations for the PMT base and two configurations for the TPB coating. All have shown very stable performance over more than 7 months of operation. All four different configurations have advantages as well as drawbacks to be taken into account in future experiments.

The accumulated data allowed to study for the first time the primary and secondary light as a function of various operation parameters in a tonne-scale experiment. The not well-known propagation mechanisms, in particular the Rayleigh scattering length and the material reflectivities, could not be extracted from our data. The impact of these parameters on the amount of collected light has nevertheless been shown. The scintillation light production processes have also been studied as a function of the electric drift field. The light loss due to recombination has been measured and is in agreement with the literature. The liquid argon purity has been monitored and was found to be very stable over time, in agreement with the electron lifetime measurement. The time profile of the S1 signal has been studied. An unexpected decrease of the slow decay time with the increase of the drift field was observed. The relative amount of light created in singlet over triplet states was measured to be increasing with the drift field. 
Using the electroluminescence light, the drift velocity of electrons at two drift fields was measured. Finally, the increase of the S2 light with the amplification field was studied. The data was also compared to detailed MC simulations, and the agreement was fairly good. 

The results presented in this paper are of great interest for the liquid argon TPC community. 
In particular, some insights of the liquid argon scintillation light production mechanisms dependence with the drift field has been shown. 
In the scope of future large LArTPC detectors like DUNE, the scintillation light will be a crucial input to enhance the calorimetry potential of the detector and to improve the discovery sensitivity to rare events searches. Therefore, these results should be further investigated in dedicated setups in order to better understand the light production and propagation mechanisms in LAr.

The operation of the \three demonstrator has been a milestone towards the design of larger dual phase LArTPCs. The design of key sub-systems, such as the extraction grid and the LEMs, has been improved based on the technical problems encountered during data taking. The analysis of the charge and scintillation light signals were used to optimize the detector configuration of the currently operating ProtoDUNE-DP. Given the comparable performances of the two PMTs base designs and the technical advantages provided by the positive base one, all PMTs installed in ProtoDUNE-DP are using this solution. Based on the \three experience, the wavelength shifter (TPB) applied directly on the PMT photocathode has been chosen as the reference design in ProtoDUNE-DP. An alternative solution, using a PEN sheet~\cite{Kuzniak:2018dcf} fixed above the PMT, has also been implemented on some PMTs. Finally, the ProtoDUNE-DP PMT layout has been driven by the Monte Carlo simulations which were validated by \three data/MC comparisons.

\acknowledgments
We are truly grateful to the strong and continuous support of CERN for the experimental infrastructure, the development of the cryogenic system and for the detector operation. We are also thankful to the CERN IT department and the IN2P3 Computing Center (CC-IN2P3) for the computing resources needed for data storage, processing, analyses and simulations. \\
This work would not have been possible without the support of the Swiss National Science Foundation, Switzerland; CEA and CNRS/IN2P3, France; KEK and the JSPS program, Japan; Ministerio de Ciencia e Innovaci\'on in Spain under grants FPA2016-77347-C2, SEV-2016-0588 and MdM-2015-0509, Comunidad de Madrid, the CERCA program of the Generalitat de Catalunya and the fellowship (LCF/BQ/DI18/11660043) from ``La Caixa'' Foundation (ID 100010434); the Programme PNCDI III, CERN-RO, under Contract 2/2020, Romania. 
This project has received funding from the European Union's Horizon 2020 Research and Innovation program under Grant Agreement no. 654168.
The authors are also grateful to the French government operated by the National Research Agency (ANR) for the LABEX Enigmass, LABEX Lyon Institute of Origins (ANR-10-LABX-0066) of the Universit\'e de Lyon for its financial support within the program "Investissements d'Avenir" (ANR-11-IDEX-0007).

\bibliographystyle{JHEP}
\bibliography{biblio}

\end{document}